\documentclass[prd,aps,tightenlines,superscriptaddress,floatfix,showpacs,
nofootinbib,eqsecnum,amsmath]{revtex4}
\usepackage{amssymb,amsfonts,latexsym, subfigure, psfrag, soul}
\usepackage{fancyhdr}
\usepackage{acronym}
\usepackage{dcolumn}

\def\RCS$#1: #2 ${\expandafter\def\csname RCS#1\endcsname{#2}}
\RCS$Date: 2006/07/18 16:01:49 $
\RCS$Revision: 1.4 $
\usepackage{graphicx}

\acrodef{LAL}{LIGO Algorithms Library}

\def\gsim{\mathrel{ \rlap{\raise 0.511ex \hbox{$>$}}{\lower 0.511ex \hbox{$\sim$}}}}
\begin{document}

\pagestyle{plain}
\fancypagestyle{plain}
\rfoot{}
\cfoot{\arabic{page} of \pageref{theend}}
\lfoot{}
\pagestyle{plain}
\lhead{Template placement -- v.\RCSRevision, \RCSDate}
\rhead{Template placement for inspiral searches}

\title{A template bank to search for gravitational waves from inspiralling
compact binaries I: physical models
}

\author{S. Babak}
\affiliation{Albert Einstein Institute, Golm, Germany}
\author{ R. Balasubramanian}
\author{ D. Churches}
\author{ T. Cokelaer}
\author{ B.S. Sathyaprakash}
\affiliation{School of Physics and Astronomy, Cardiff University, Cardiff CF24
3YB, UK}

\begin{abstract}
Gravitational waves from coalescing compact binaries are searched using the
matched filtering technique. As the model waveform depends on a number of
parameters, it is necessary to filter the data through a template bank
covering the astrophysically interesting region of the parameter space. The
choice of templates is defined by the maximum allowed drop in signal-to-noise 
ratio due to the discreteness of the template bank.  In this paper we describe 
the template-bank algorithm that was used in the analysis of data from the 
Laser Interferometer Gravitational Wave Observatory (LIGO) and GEO\,600 detectors to 
search for signals from binaries consisting of non-spinning compact objects. 
Using Monte-Carlo simulations, we study the efficiency of the 
bank and show that its performance is satisfactory for the 
design sensitivity curves of ground-based interferometric gravitational 
wave detectors GEO\,600, initial LIGO, advanced LIGO and Virgo.
The bank is efficient to search for various compact binaries such 
as binary primordial black holes, binary neutron stars, binary
black holes, as well as a mixed binary consisting of a non-spinning black hole
and a neutron star.
\end{abstract}

\maketitle

\section{Introduction}

Long baseline interferometric gravitational wave detectors are either 
already taking sensitive data (LIGO \cite{LIGO}, GEO\,600 \cite{GEO600} and TAMA
\cite{TAMA}) or will be ready to do so in the near future (Virgo \cite{VIRGO}).
The most promising candidates for these detectors are coalescing binaries
consisting of compact objects such as neutron stars (NS) and/or black holes (BH)
in close orbit, losing energy and angular momentum through the emission 
of gravitational radiation.

In the past decade a lot of research has been carried out in modelling 
the late-time dynamics of compact binaries and the emitted radiation. 
To this end post-Newtonian (hereafter PN) approximation has been used in which
all the relevant physical quantities are expressed as a perturbative 
series in the parameter $v$ which is a measure of the relative 
velocity of the two component masses of the binary.
The approximation is now available to a rather high order $v^7$ in PN
theory \cite{BDIWW,BDI95,BIWW96,BFIJ,LB,BDEI04}. These models are definitely accurate
enough to serve as faithful templates for matched filtering the radiation from
binary neutron stars (hereafter, BNS) and binaries consisting of primordial
black holes (hereafter, PBHs) \cite{NakamuraEtAl}. 
This is because both BNS and binary PBHs
merge outside the sensitive band of current ground-based instruments and
strong relativistic effects that come into play at the time of merger but
not properly modelled by the PN theory are unimportant for their 
observation.  In the case of binary black holes (BBH), however, PN
models are not quite accurate since these systems will be ultra-relativistic 
($v\sim 0.2$-$0.4$) when their frequencies enter the band where the detectors
are most sensitive and some of the BBH systems would even coalesce at
frequencies where the detectors have the best sensitivity. There are now models
that go beyond the PN theory to predict the waveform given out by these systems
and the different models for the binary dynamics predict significantly different
waveforms. 

\subsection{Signal models used in the search}

Current searches for gravitational waves from inspiralling compact binaries 
employ template waveforms that are based on the {\it adiabatic, 
non-adiabatic} and {\it phenomenological} models. The
adiabatic waveforms are obtained by using the PN expansions of the
conserved energy $E$ and flux $\cal F$ and solving the energy balance equation
$dE/dt = -\cal F,$ which in turn leads to an evolution equation for
the angular velocity $\omega(t)$ and hence the phasing of gravitational waves
$\varphi(t) = (2\pi)^{-1} \int \omega(t) dt.$ An implicit assumption
made in writing the energy balance equation is that $E$ does not change (appreciably)
over an orbital timescale. There has been significant amount of activity to
understand the poor behaviour  \cite{Poisson,DIS1} of the PN scheme 
resulting in new improved models that are based on the use of re-summation
methods to accelerate the convergence of the PN expansions.
The Pad\'e re-summation method \cite{DIS1} uses rational polynomial approximations of the
original Taylor expansions to improve the convergence of the PN
series. The non-adiabatic models directly integrate the
equations of motion (as opposed to using the energy balance equation) and there is no
implicit conservation of energy that is used in the orbital dynamics as, for instance, in the 
effective-one-body (EOB) approach 
\cite{EOB1,EOB2,EOB3,DIS2}.  The EOB maps the real two-body conservative 
dynamics onto an effective one-body problem wherein a test
mass moves in an effective background metric.

The PN, Pad\'e and EOB models produce physical waveforms, as opposed to the
phenomenological waveforms \cite{BCV1,phenomenological templates,Kalogera}. The latter waveforms capture 
the main features of the physical waveforms but use unphysical parameters.
Nevertheless the phenomenological waveforms themselves fit very well with both
the adiabatic and non-adiabatic models, because the scheme uses greater number of
free parameters and the parameters are extended to vary over a wider range than
that which is physically allowed.

\subsection {Matched filtering and template banks}

Generally speaking, since the signal from a coalescing binary can be modelled
reasonably well we can employ matched filtering, which is the optimal strategy
to search for signals  buried in stationary, Gaussian noise.  However,
real interferometer noise is neither stationary nor Gaussian, and we have to
use additional consistency checks, such as the $\chi^2$-test 
\cite{Ballen,Babak}, to discriminate true events from broadband 
transients of instrumental or environmental origin.

The matched filtering performs a phase-coherent correlation of the data with a
template which is a copy of the model waveform weighted down by the expected 
power spectral density of the data. The 
gravitational wave signal
depends on a set of continuous parameters such as the amplitude, the masses
$m_1,\, m_2,$ and spins of the stars and a fiducial reference time (taken to be
either ``time-of-arrival'' $t_0$ or ``epoch-of-coalescence'' $t_C$) and the corresponding
orbital phase $\phi_0$. The parameters describing the inclination of the binary orbit 
and the location of the source on the sky are degenerate with other parameters,
and are therefore, ignored in the search problem, although they can be 
deduced once the response of three or more non-co-located detectors to 
a gravitational wave event
is known. The eccentricity may also be ignored, since it is expected to be
negligible (due to orbital circularization) by the time the gravitational
radiation enters the frequency band (10-1000 Hz) of a ground-based detector. In this
 paper we will only concentrate on non-spinning systems for which the dimension of the
 search is restricted to 4 parameters ($m_1,\, m_2,$, $t_C$ and $\phi_0$ ) excluding 
the additional 8 parameters to be used in a spinning system search (angular
momentum of the system and the spin orientation of each object).

Since the parameters of an incoming gravitational wave signal 
will not be known a priori, the data must be searched with 
a set of templates, called {\it a
template bank}, designed to cover the whole parameter space.  The
distance between the templates in the parameter space is governed by the
trade-off between computational power and loss in detection rate due to
the discrete nature of the template bank. The spacing should be chosen so that
the loss in signal-to-noise ratio due to the mis-match of the template with
the signal does not deprecate the detectability of these sources. 

We employ a geometrical approach 
\cite{SD91,DS92,BSD95,Owen96,OwenSathyaprakash98}
in constructing a template bank, namely we
define a metric on the parameter space and use local flatness theorem to place
templates at equal distances. The template bank could be seen naively as a uniform grid in the
parameter space. However, to emphasize that this is not, in general a trivial task, we
suggest to consider placing dots on a sphere at equal distances. Placing
templates on the parameter space depends upon two crucial things: (i)
geometrical properties of the signal manifold, whether it is flat or curved, and
(ii) the coordinates chosen for template placement, which could be curvilinear
even when the manifold is flat, or ``almost flat''. Indeed, we need to choose a
``Lorentzian-like'' (or close to it) coordinate frame. The second point is
similar to the problem of laying a regular grid on the plane using either
Cartesian or polar coordinates. The situation is even more complicated because
the metric, in general, depends on the waveform model and the PN
order one chooses. 

\subsection {Filtering with physical and phenomenological templates}

The different PN models differ significantly from each other in their 
prediction of the phasing of the waves, making it necessary to search
for binary black holes and neutron stars by using not one but all
the different families of waveforms currently available.
Luckily enough, as we will show in this paper and an accompanying paper
\cite{BCS2}, it turns out that we can use a single bank for {\it all} the
different physical waveforms explored thus far. This bank is currently used by
the GEO\,600 and LIGO detectors to search for BNS, binary PBHs and
BBH \cite{LIGOS2bns,LIGOS2bbh,LIGOS2PBH}. 
The signal from a BNS sweeps through the detector bandwidth and
the two neutron stars can even merge outside the sensitive bandwidth of the
current ground-based detectors and the waveforms corresponding to
different physical models agree reasonably well among each other (with overlaps very close to unity). For higher mass systems, e.g. BBH, 
the situation changes, the
deviation between the various models getting more significant in the sensitive
bandwidth of the detectors.

Since for BBH the model waveforms from different families do not agree well with
each other, a template bank that employs one PN family of templates will not be
efficient to detect signals from another family. One solution is to filter the
data through a number of banks with templates based on all the different models.
If the reasonable assumption that at least one of the available PN models is
close to the true gravitational wave signal bears out then our bank consisting
of templates from all the different PN families should be efficient enough in
capturing the coalescence events. Unexpectedly, we have found that the bank
works well as long as the template and signal are both from the same family
meaning a single template bank might suffice to filter the data through
different PN families.

Another approach to this problem is to use a single template bank of
phenomenological waveforms \cite{BCV1}. Buonanno, Chen and Vallisneri
have designed phenomenological waveforms that are motivated by 
the structure of the PN signal in the Fourier domain which, for high
mass systems, has good overlaps with all the known models. They are
able to do this by (a) introducing (in the case of non-spinning binaries)
two more parameters than are necessary in the case of physical waveforms
and (b) allowing the parameters to vary over a wider, and in some
cases unphysical, range than in the case of physical waveforms.
The advantage in using the
phenomenological templates is that they have been designed to give a good fit
not only to known models but also to waveforms which could be ``in between'' these models.  
However, there is also a disadvantage in using these waveforms as they are very
poor in estimating the physical parameters of the emitting system. Moreover,
using two filters instead of one increases the computational cost and the
threshold to be used. It produces more false alarm due to larger size of 
the corresponding template bank. So we might want to filter the data with 
candidate events through a bank of {\it physical} 
templates for the sake of parameter estimation.

In a companion paper we shall define and study the efficiency of a phenomenological
template bank to capture different physical models of gravitational wave
signals using designed and real sensitivity curves of the various interferometers. We
have verified that such a template bank performs well with the minimal expected (or
higher) efficiency for signals from BBH systems with total mass in the
range $6$-$40\,M_\odot$ for designed sensitivity.
For sensitivity seen during the second Science run of the LIGO interferometers
the bank covers the parameter space with good efficiency in the same range of
the total mass and a lower frequency cutoff of 40 to 100 Hz.

\subsection{Organisation of the paper}

In the rest of this paper we will discuss only issues related to the physical
template bank where both signals and templates are based on the 
physical models of the gravitational waveform from an inspiralling binary. 
In Sec.~\ref{sec:spa} and \ref{sec:filtering} we
briefly review, respectively, the Fourier-domain representation of the signal  
and the technique of matched filtering in a
geometrical language, with the aim of introducing the notation and language
used in Section \ref{sec:template placement} to construct a template bank.
Section~\ref{sec:template placement}
constructs a generic template bank to search for binary inspirals from sources
with a wide range of masses and applicable to any interferometer.  The template
bank specifically targets equal mass binaries but it is suitable for binaries
with a small mass ratio as long as the precessional effects due to the spin of
the larger body are unimportant. Section~\ref{sec:efficiency} discusses the
results of Monte-Carlo simulations performed with different design sensitivity
curves from a variety of interferometers. Section~\ref{sec:conclusions}
summarizes the main results of this paper and discusses the application of
the template bank introduced in this paper to capture other target models to
be presented in a separate publication.

\section{Inspiral signal in the stationary phase approximation}
\label{sec:spa}
For the purpose of the template bank design it suffices to use a simple
model of the signal. We shall use the Fourier representation of the
standard PN waveform in the so-called restricted PN
approximation. In this approximation, one neglects the PN
corrections to the amplitude of the signal while the corrections to the phase
are fully taken into account to the highest order possible/available. 

Let us begin with the time-domain representation of the waveform. 
The response of an interferometric gravitational wave detector to arbitrarily 
polarized waves from an inspiralling binary of total mass $M=m_1+m_2,$ 
mass ratio $\eta=m_1 m_2/M^2$, at a distance $D$ is given by 
\begin{equation}
h(t) = \frac {4A\eta M}{D} \left [\pi M f(t)\right ]^{2/3} 
\cos [\varphi(t) + \varphi_C],
\label{eq:waveform1}
\end{equation}
where it is assumed that the detector's motion is unimportant during the
time when the signal sweeps across its bandwidth. Here,
$f(t)$ is the (invariant) instantaneous frequency of the signal
measured by a remote observer, the phase of the signal
$\varphi(t)=2\pi\int^{t_C} f(t) dt$ is defined so that it is zero when the
binary coalesces at time $t=t_C,$ $\varphi_C$ is the phase of the signal
at $t_C$ and $A$ is a numerical constant whose value depends on the
relative orientations of the interferometer and the binary orbit which when
averaged over all angles is 2/5 \cite{DIS2}.

One can compute the Fourier transform $\tilde{h}(f)$ of the waveform given in
Eq.~(\ref{eq:waveform1}) using the stationary phase approximation \cite{BDI95,BSD95}:
\begin{eqnarray}
\label{eq:SPA}
\tilde{h}(f) & = & \frac {A M^{5/6}}{D\pi^{2/3}}
\sqrt{\frac{5\eta}{24}} f^{-7/6} 
\exp\left [ i \Psi(f;\,t_C,\,\varphi_C,\,\lambda_k) + i \frac{\pi}{4}\right ]\\
\Psi(f) & = &  2\pi f t_C + \varphi_C + \sum_k \lambda_k f^{(k-5)/3}.
\label{eq:SPA phase}
\end{eqnarray}
The parameters $\varphi_C$ and $t_C$ are the so-called {\it extrinsic,} or {\em
kinematic}, parameters and they are defined by the relative orientation and
position of the source and the detector and a fiducial time-of-coalescence of
the binary. Intrinsic parameters $\lambda_k$ are the {\it chirp parameters}  related to the two
masses of the binary. At the 2PN order there are four non-zero 
{\it chirp parameters}, $\lambda_0, \lambda_2,\lambda_3, \lambda_4$ ($\lambda_1=0$
owing to the fact that there is no 0.5PN term in the PN expansion), given by
\begin{eqnarray}
\lambda_0 & = & \frac{3}{128\eta (\pi M)^{5/3}}, \  \
\lambda_2 = \frac{5}{96 \pi\eta M}   
\left ( \frac{743}{336} + \frac{11}{4} \eta\right ),\ \  
\lambda_3 = \frac{-3\pi^{1/3}}{8\eta M^{2/3}},   \nonumber\\ 
\lambda_4 & = & \frac{15}{64\eta (\pi M)^{1/3}}   
\left ( \frac{3058673}{1016064} + \frac{5429}{1008} \eta +   
\frac{617}{144}\eta^2 \right).
\label {eq:chirp parameters}
\end{eqnarray}   
The chirp parameters are all functions of the two mass parameters -- the total
mass $M$ and the symmetric mass ratio $\eta$. Instead of the masses we can
choose any two of the chirp parameters to characterize the signal. The choice
$(\lambda_0,\, \lambda_3),$ in particular, is convenient since the masses 
can be computed explicitly in terms of these two parameters:
\begin{equation}
M = \frac{-\lambda_3}{16 \pi^2 \lambda_0}, \ \ 
\eta = \frac{3}{4\lambda_0} 
\left (\frac{-2\pi \lambda_0}{\lambda_3} \right )^{5/3}.
\label{eq:masses in chirp parameters}
\end{equation}
Often, it is convenient to use chirptimes $\tau_0$ and $\tau_3,$ in lieu of 
chirp parameters $\lambda_k,$ given by 
\begin{equation}
\tau_0 = \frac{5}{256 \pi f_L \eta} \left(\pi  M f_L\right)^{-5/3}, \ \ 
\tau_3 = \frac{1}{8  f_L \eta}  \left( \pi  M  f_L\right)^{-2/3},
\label{eq:chirp parameters in masses}
\end{equation}
where $f_L$ is the lower-cutoff
frequency at which the generation of the templates starts. The value assigned to
$f_L$ should be chosen carefully: while a low value increases the size of 
the template bank a high value decreases the signal-to-noise
ratio. Section~\ref{subsec:fl} discusses how to optimally determine 
the lower frequency cutoff. Finally, the chirptimes can also be inverted 
in terms of the masses \cite{Mohanty}:
\begin{equation}
M = \frac{5}{32 \pi^2 f_L} \frac{\tau_3}{\tau_0}, \ \ 
\eta = \frac{1}{8 \pi f_L \tau_3} 
\left (\frac{32\pi \tau_0}{5 \tau_3} \right )^{2/3}.
\label{eq:masses in chirp times}
\end{equation}
The correspondence between the two sets of parameters  $(m_1\,, m_2)$ and
$(\tau_0,\, \tau_3)$  is illustrated in Fig.~\ref{fig:t0t3}

\section{Matched filtering search for inspirals}

\label{sec:filtering}

A matched filter is an optimal linear filter to detect a signal of known shape
in stationary, Gaussian noise. It can be introduced in several ways and here we
shall follow the geometrical formalism \cite{BSD95,Owen96} of signal analysis which is most
convenient for the construction of a template bank \cite{Owen96}. The geometrical picture
is more easily motivated by considering the detector output as a discrete
time-series rather than a continuous function. Without loss of generality
we shall follow this standard procedure in summarizing the basic results
from differential geometric approach to signal analysis but often we shall 
also write the equivalent formulas obtained in the continuum limit.
Although the signal is considered as a discrete series in time, it will be treated
as a continuous function of the signal parameters\footnote{The discussion
in this Sec. is quite generic. Thus, we shall use the symbol $\vartheta$ to
denote all the parameters of the signal, both intrinsic and extrinsic. In
specific applications it might be easier to treat the maximisation of the 
signal-to-noise ratio 
over some of the extrinsic parameters as we shall briefly discuss in the
beginning of the next Section.}, which we shall denote as $\vartheta^\mu.$
The notation and language followed here closely resembles that in
Refs.\, \cite{CutlerAndFlanagan,Owen96,OwenSathyaprakash98}. 

\subsection{A Geometric approach to signal analysis}
The output $x(t)$ of a detector is not recorded continuously;
it is sampled discretely, usually at equal intervals $\Delta t,$
chosen so as to obey the sampling theorem 
(see e.g., Ref.\ \cite{PercivalWaldon}).
Thus, the detector output lasting for a time $T$ and sampled at a rate $f_s$
($f_s\equiv (\Delta t)^{-1}$) is a time-series consisting of
$N$-samples, $N=T/\Delta t.$ The set of all possible detector outputs forms
an $N$-dimensional vector space. A gravitational wave signal characterized by
a set of $p$ parameters $\vartheta^\mu,$ $\mu=0, 1, \ldots, p-1,$ sampled in
a similar way is also an $N$-dimensional vector. 
However, the set of all signal vectors {\it
don't} form a vector space, although they {\it do}
form a manifold (a $p$-dimensional surface in an $N$-dimensional vector space), 
with the parameters serving as a coordinate system. 

Further structure can be developed on this
manifold by defining a metric. It turns out that the statistic of matched
filtering naturally leads to the definition of a scalar product between vectors,
which can then be used to induce a metric on the signal manifold. We shall now
turn to a brief discussion of the matched filtering in geometrical language.

Let $x_k = n_k + s_k$ [$x(t)=n(t) + s(t)$] denote the output of a 
stationary\footnote{Stationarity is the statement that the
statistical properties of a detector, i.e. the distribution
of its output, remain the same as a function of time. 
Mathematically, one expresses stationarity by demanding
that the correlation of the noise at two different 
instants depends only on the absolute time-difference
of the two instants but not on the two instants.} 
detector consisting of instrumental 
and environmental noise background $n_k$ [$n(t)$] and a possible 
deterministic signal $s_k$ [$s(t)$]. 
We treat $n_k$ as a random time-series 
drawn from a large ensemble whose 
statistical properties are those of the detector noise. 
We use $\left <n_k \right >$ to denote the average over 
the ensemble and assume that $n_k$ is
a stationary Gaussian random process with zero mean.
In the frequency domain we denote these quantities
as : $\tilde{x}_m = \tilde{n}_m + \tilde{s}_m$ [$\tilde{x}(f)=
\tilde{n}(f) + \tilde{s}(f)$], where 
Fourier series $\tilde{s}_m$ [$\tilde{s}(f)$] 
of the time series $s_k$ [$s(t)$] is defined by 
\begin{equation}
\tilde{s}_m = \sum_{k=0}^{N-1} s_k e^{-2\pi i m k/N},\ \ \ \ 
\tilde{s}(f) = \int_{-\infty}^\infty s(t) e^{-2\pi i f t} dt.
\end{equation} 

Next,  let us introduce the Hermitian inner product of vectors.
Given vectors $g$ and $h$ their inner product, denoted $(g,h),$
is defined as 
\begin{equation}
(g,h) = 2 \sum_{k=0}^{N-1}
\frac{\tilde{g}_k^*\tilde{h}_k + \tilde{g}_k\tilde{h}^*_k}{S_{hk}},\ \ \ \ 
(g,h) = 2 \int_0^\infty
\frac{\tilde{g}(f)^*\tilde{h}(f) + \tilde{g}(f)\tilde{h}(f)^*}{S_h(f)} df.
\label{inprod}
\end{equation}
Here ``$*$'' denotes the complex conjugate and $S_{h}$ is the 
one-sided noise power spectral density (PSD) given by
\begin{equation}
\left <\tilde{n}_k\tilde{n}^*_m\right > = \frac{1}{2}S_{hk}\delta_{km},\ \ \ \
\left <\tilde{n}(f_1)\tilde{n}^*(f_2)\right > = \frac{1}{2}S_h(f_1)
\delta(f_1-f_2).
\label{eq:noise psd}
\end{equation}
The inner product naturally leads to the concept of 
the norm $||h||$ of a vector $h:$ $||h|| = \sqrt{(h,h)}$. 
A vector $h$ is said to be normalized if its norm is equal
to unity. If the vector $h$ is to begin with not normalized then we can
define a vector with unit norm by
\begin{equation}
 \hat{h} = \frac{h}{\sqrt{(h,h)}}.
\end{equation}

It follows from the theory of hypothesis testing that in order
to detect a {\it known} signal,  but with an {\it unknown} set of
parameters $\vartheta^\mu,$ in stationary Gaussian background noise $n$
one must construct the statistic $\xi$ given by
\begin{equation}
\xi = 2 \sum_{k=0}^{N-1}
\frac{\tilde{x}_k^*\tilde{h}_k(\vartheta^\mu) + 
\tilde{x}_k\tilde{h}^*_k(\vartheta^\mu)}{S_{hk}} = 
\left (x, h(\vartheta^\mu) \right )
\label{eq:mf statistic}
\end{equation}
and maximize it over the parameters $\vartheta^\mu.$ Here $h$ is the
template and $q \equiv h/S_h$ is the optimal filter.  
However, the terms template and optimal filter are 
often used interchangeably to mean either of $h$ or $q.$ 
Writing the above equation in terms of $q$ we see that the 
statistic $\xi$ is the cross-correlation of 
the detector output with the optimal filter $q$ and not 
just with the template $h.$ Although at first this seems
strange, upon closer inspection one can see why the optimal
filter is not just a copy of the signal we are looking
for but the signal divided by the noise PSD. The noise PSD
weighs the correlation differently at different frequencies,
making greater contribution to the
region of the spectrum where the sensitivity is larger
(i.e., $S_h$ is smaller) and smaller contribution where
the sensitivity is weaker (i.e., $S_h$ is larger). 

The signal-to-noise ratio (SNR) $\rho$ is the mean of the
statistic $\xi$ divided by the square-root of its variance given by
\begin{equation}
\rho = \frac{\left< \xi \right >}
{\sqrt{\left <\xi^2\right > - \left <\xi \right >^2}}
= \frac{\left <(x,h)\right >}
{\sqrt{\left <(x,h)^2\right > - \left <(x,h)\right >^2}} = 
\frac{(s,h)} {\sqrt{(h,h)}},
\end{equation}
where the last equality follows from Eq.~(\ref{eq:noise psd})
and our assumption that the noise is 
a stationary random process with zero mean.  Thus, the effect of filtering
a signal $s$ with a template $h$ is equivalent to projecting the signal
$s$ in the direction of $h,$ i.e. $\rho=(s,\hat h).$ 
This projection is, obviously, the greatest
when $h=s$ (the matched filter theorem), giving
$\rho = \sqrt{(s,s)}.$ When the template is not the same
as the signal the SNR is less than the optimal value. That is,
in general $\rho \le \sqrt{(s,s)}.$

\subsection{The mis-match metric}
\label{sec:metric}
We can use the inner product Eq.~(\ref{inprod}) to induce a
metric on the signal manifold.  In the rest of this Section 
we shall assume that our templates are normalized, that is $(h,h)=1.$
The distance between two infinitesimally separated normalized 
templates on the signal manifold is given by
\begin{eqnarray}
||h(\vartheta^{\mu} + d\vartheta^{\mu}) - h(\vartheta^{\mu}) ||^2
& = &  \left|\left| h_{\mu} d\vartheta^{\mu} \right|\right|^2 
       \nonumber \\
& = & \left( h_{\mu}, h_{\nu}\right)\,d\vartheta^{\mu}
      d\vartheta^{\nu} \nonumber \\
& \equiv & g_{\mu\nu}d\vartheta^{\mu}
      d\vartheta^{\nu},\label{norm}
\end{eqnarray}
where $h_\mu$ is the partial derivative of the signal $h$ with
respect to the parameter $\vartheta^\mu$ and Einstein's convention
of summation over repeated indices is assumed. The quadratic form 
$g_{\mu\nu} \equiv (h_\mu,\, h_\nu)$ 
defines the metric induced on the 
signal manifold. This metric is the same as the 
{\it mismatch} metric defined in Ref.\ \cite{Owen96,OwenSathyaprakash98}.
Let $\mathcal{O}(\vartheta^\mu,\vartheta^\mu+\delta\vartheta^\mu)$ 
denote the overlap between two infinitesimally separated vectors
$h(\vartheta^\mu)$ and $h(\vartheta^\mu+\delta\vartheta^\mu),$ that is
\begin{equation}
\mathcal{O}(\vartheta^{\mu}, \vartheta^{\mu}+\delta\vartheta^{\mu})  
\equiv (h(\vartheta^{\mu} + \delta\vartheta^{\mu}), 
  h(\vartheta^{\mu})).
\end{equation}
Since our templates are normalized we have 
$\mathcal{O}(\vartheta^\mu,\vartheta^\mu)=1.$ For vectors that are
infinitesimally close to each other we can expand the overlap about
its maximum value of 1 ($\delta\vartheta^{\mu}\rightarrow0$):
\begin{equation}\label{eq:overlap}
\mathcal{O}(\vartheta^{\mu}, \vartheta^{\mu}+d\vartheta^{\mu}) 
= 1 - \left.M_{\mu\nu}\right|_{\vartheta^{\alpha}}
\delta\vartheta^{\mu}\delta\vartheta^{\nu} + \ldots,
\end{equation}
where $M_{\mu\nu}=-\frac{1}{2}\frac{\partial^2 {\cal O}}
{\partial\vartheta^\mu \partial\vartheta^\nu}.$
The {\it mismatch} $M$ between the two vectors is 
$M \equiv 1-\mathcal{O}(\vartheta^{\mu}, \vartheta^{\mu}+d\vartheta^{\mu}) 
= g_{\mu\nu}\delta\vartheta^{\mu}\delta\vartheta^{\nu}.$
Thus, the metric on the manifold can be equivalently constructed
using either of the following formulas \cite{Porter}:
\begin{eqnarray}
g_{\mu\nu} & = & (h_\mu, h_\nu) \label{eq:metric1} \\
           & = & -\frac{1}{2}\frac{\partial^2 {\cal O}}
                 {\partial\vartheta^\mu \partial\vartheta^\nu}.
\label{eq:metric2}
\end{eqnarray}

Next, we introduce the concept of the {\it minimal match} 
$MM$ \cite{SD91,Owen96,OwenSathyaprakash98}. In searching for
a signal of known shape but unknown parameters one filters
the data through a bank of templates. The templates in the bank
are copies of the signal corresponding to a set of values 
$\vartheta_i^\mu,$ $i=1,2\,\ldots$. The parameters
$\vartheta_i^\mu$ are chosen 
in such a manner that (a)~the bank of templates has at least 
an overlap of $MM$ with any signal whose parameters are in 
a given range
\begin{eqnarray}
\min_{\vartheta^{\mu}} \max_{\vartheta^{\mu}_i}
(h(\vartheta^{\mu}), h(\vartheta^{\mu}_i)) \ge MM,
\end{eqnarray}
and (b)~the number of templates is the smallest for the chosen 
value of $MM.$ The quantity $MM$ is a very important notion in constructing a bank of templates. 
Indeed, if we assume that the incoming 
gravitational wave signal $s(\vartheta^{\nu})$ is reproduced exactly 
by our model, so that $s(\vartheta^{\nu}) = A \hat{h}(\vartheta^{\nu})$, 
then we could have a drop in the SNR due to the coarseness of the 
template bank. Indeed, if the template nearest to the signal is
$\hat{h}(\vartheta^{\nu}_i)$ then 
the SNR is given by
\begin{equation}
\rho = \max_{h_i}(s,\hat{h_i}) = A \mathcal{O}(\vartheta^{\mu}, 
\vartheta^{\mu}_i) \ge A \times MM.
\end{equation}
In reality, however, the overlap between the template and the
signal can be smaller than the 
minimal match since our template may not be an accurate representation
of the signal.  Recall that our templates are based on the solution
to a PN expansion of the Einstein's equations.
Thus, our template might not be sufficiently accurate in describing the 
fully general relativistic gravitational wave signal emitted by a binary 
during the final stages of the coalescence of
two compact objects. A lot of effort in the last 15 years has been focused on
improving the PN templates by (a)~working out the expansion to
higher orders, and (b)~employing re-summation
techniques that accelerate the convergence of the PN expansion.
Current searches for gravitational waves deploy several families
of template banks representing the different classes of the signal models.
If the signal $s$ and the template $h$ are from different families then an
important notion that will be useful in studying the efficiency of
the template bank is the so-called {\it fitting factor} introduced in
Ref.~\cite{Apostolatos} and defined as:
\begin{equation}
FF(s,h(\vartheta^{\mu}_i)) = \max_{\vartheta^{\mu}_i} (\hat{s}, \hat{h}(\vartheta^{\mu}_i))
\label{FF}
\end{equation}
The fitting factor incorporates the degree of faithfulness of our templates in
detecting the expected signal $s$.

Finally, given the minimal match and the metric on the parameter space one can
estimate the number of templates required to cover the desired range
of parameters using \cite{Owen96}
\begin{equation}
N_b[g,MM] = \left( 2\sqrt{\frac{1-MM}{p}} \right)^{-p}
\int \sqrt{|g|}d\vartheta^{\mu},
\label{numTemp}
\end{equation}
where $p$ is the number of independent parameters. 

Strictly speaking, thequadratic approximation used in Eq.~\ref{eq:overlap} is
valid for $\delta\vartheta^{\mu}\rightarrow0$ \cite{Owen96} but for practical
purposes the approximation is valid for $MM=90\%-95\%$. 

\subsection{Application to binary inspirals}
\label{sec:appln to inspirals}
Let us now apply the geometrical language of the previous
Section to the case of binary inspirals. Our aim is to derive an explicit
expression for the metric and then use it to set up a bank of templates.
The discussion in the previous Section illustrated the general method
for obtaining the metric on a signal manifold whose dimension is
equal to the number of parameters characterising the signal.
It might at first, therefore, seem as though we would need templates
in the full multi-dimensional signal parameter space. 
We need a bank of templates since we would
not know before hand what the parameters of the signal are and have to
construct the detection statistic for different possible values of the
signal parameters, such that at least one template in our bank is
close enough to any possible signal so as not to lose more than, say
5\% of the SNR. However, templates are not always explicitly needed
to maximize the SNR (as shown by Schutz \cite{SchutzInBlair})
in the case of the extrinsic
parameters $t_C$ and $\varphi_C,$ thereby greatly reducing the 
problem of having to filter the data through a large template bank.
Indeed, data analysts are always looking for ways to reduce the
effective dimensionality of the template space. 

The maximization over the phase $\varphi_C$ is
achieved via the decomposition of the signal into its quadratures: two templates
with a $\pi/2$ phase difference, e.g. $\varphi_C=0$ and $\varphi_C=\pi/2$
\cite{SchutzInBlair,SD91}
\begin{equation}
\max_{\varphi_C} \rho = \max_{\varphi_C} (x,\hat{h}_{i}(\varphi_C)) =
\sqrt{(x,\hat{h}_i(0))^2 + (x,\hat{h}_i(\pi/2))^2}.
\end{equation} 
Thus, we need only two templates to search in the $\varphi$-dimension.
Similarly, the search for the fiducial time $t_C$ at which the binary 
merges can be efficiently found in practice via the fast Fourier 
transform
\begin{equation}
\max_{t_C} \rho = \max_{t_C}\rho(t_C),\ \ \ \  
\rho(t_C) = \sum_m \frac{\tilde{x}^*_m 
\hat{h'}_m}{S_{hm}}e^{2\pi i m t_C}.
\end{equation}
Thus, we need the metric and the templates only in the two-dimensional 
space of intrinsic parameters, which are taken to be the chirp 
parameters $\tau_0$ and $\tau_3$ instead of the masses $m_1$ 
and $m_2$ of the binary. However, it is more convenient to begin
with the metric in the three-dimensional space of $(t_C, \tau_0,
\tau_3)$ and then project out the coordinate $t_C.$

In practice, we can also work with the chirp parameters
$\theta_1$ and $\theta_2$ :
\begin{equation}
\theta_1 = 2\pi f_L \tau_0,\ \ \ \ 
\theta_2 = 2\pi f_L \tau_3.
\label{eq:theta parameters}
\end{equation}

We can use either of Eqs.(\ref{eq:metric1}) or (\ref{eq:metric2}) to
work out the metric.  Owen \cite{Owen96} used Eq.~(\ref{eq:metric2})
and maximized the metric over the phase $\varphi_C$ to obtain an
explicit expression for the metric $\gamma_{\alpha\beta}$ 
in the three-dimensional space of $(t_C, \theta_1, \theta_2):$ 
\begin{equation}
\gamma_{\alpha \beta} = \frac{1}{2} \left( \mathcal{J} [ \psi_{\alpha}
\psi_{\beta} ] -
\mathcal{J} [ \psi_{\alpha} ] \mathcal{J} [ \psi_{\beta} ] \right),
\label{eq:gamma metric}
\end{equation}
where $\psi_\alpha$ is the derivative of the Fourier phase of 
the inspiral waveform with respect to the parameter $\theta^\alpha,$ 
that is $\psi_\alpha \equiv \partial \Psi/\partial \theta_\alpha,$ and ${\mathcal J}$ is
the moment functional \cite{PoissonAndWill} of the noise PSD.  For any function 
$a(x)$ the moment functional $\mathcal{J}$ is defined as,
\begin{equation}
\mathcal{J} [a] \equiv \frac{1}{I(7)} 
\int^{x_{\rm U}}_{x_{\rm L}} \frac{a(x)\, x^{-7/3}}{S_{h}(x)}\, dx.
\label{eq:moment functionals}
\end{equation}
where $I(q)$ is the $q$th moment of the noise PSD defined by, 
\begin{equation}
I(q) \equiv S_{h}(f_{0}) \int^{x_{\rm U}}_{x_{\rm L}} 
\frac{x^{-q/3}}{S_{h}(x)} \, dx.
\label{eq:moments}
\end{equation}
Here $x\equiv f/f_0$ (also $x_{\rm U}\equiv f_{\rm U}/f_0$
and $x_{\rm L}\equiv f_{\rm L}/f_0$), $f_0$ is a fiducial frequency 
chosen to control the range of numerical values of the functions entering the
integrals, $f_{\rm L}>0$ is the lower-cutoff chosen so that the contribution to
the integrals from frequencies smaller than $f_{\rm L}$ is negligible and
$f_{\rm U}$ is the upper frequency cutoff corresponding to the system's last
stable orbit frequency.  
The metric $g_{mk}$ on the subspace of just the masses (equivalently, 
chirp times) is given by projecting out the coalescence time $t_C$:
\begin{equation}
g_{mk} = \gamma_{mk} - \frac{\gamma_{0m} \gamma_{0k}}{\gamma_{00}}.
\end{equation}

We shall now derive an explicit formula for the metric on the sub-space
of masses by using the chirp parameters $\theta_1$ and $\theta_2$ from
Eq.~(\ref{eq:theta parameters}) as our coordinates.

The starting point of our derivation is the Fourier domain phase 
Eq.~(\ref{eq:SPA phase}). By writing the $\lambda_k$'s in
terms of $\theta_1$ and $\theta_2$ we get 
\begin{eqnarray}
\Psi(f; t_C, \theta_1, \theta_2) & = & 2\pi f t_C + a_{01}\theta_1 x^{-5/3}  
+ \left [a_{21} \frac {\theta_1}{\theta_2} + a_{22} \left ( \theta_1 \theta_2^2
\right )^{1/3} \right ] x^{-1}
+ a_{31} \theta_2 x^{-2/3} \nonumber \\
& + & \left [a_{41} \frac {\theta_1}{\theta_2^2} + a_{42} \left ( \frac
{\theta_1}{\theta_2} \right )^{1/3} 
+ a_{43} \left ( \frac{\theta_2^4}{\theta_1} \right )^{1/3} \right ] x^{-1/3},
\end{eqnarray}
where the constants $a_{km}$ are given by:
\begin{eqnarray}
a_{01} = \frac{3}{5}, \ \ a_{21} = \frac{11\pi}{12}, \ \ 
a_{22} = \frac{743}{2016} \left ( \frac {25}{2\pi^2} \right )^{1/3}, \ \ a_{31}
= -\frac{3}{2}, \nonumber \\
a_{41} = \frac {617}{384} \pi^2, \ \ a_{42} = \frac{5429}{5376} \left ( \frac{25
\pi}{2} \right )^{1/3},\ \ 
a_{43} = \frac {15293365}{10838016} \left ( \frac{5}{4\pi^4} \right )^{1/3}.
\end{eqnarray}
The gradients $\psi_0=\partial \Psi/\partial t_C,$ $\psi_m = \partial \Psi/
\partial\theta_m$ enter the moment functionals which can 
be computed from the foregoing expression for the Fourier phase
\begin{equation}
\psi_{0} = 2 \pi f,\ \ \ \ 
\psi_m = \sum_{k=0}^n \Psi_{mk} x^{(k-5)/3},
\end{equation}
where $n$ is the PN order up to which the phase is known, or the
PN order at which the metric is desired.
Expansion coefficients $\Psi_{mk}$ can be considered to be a 
$(2\times n)$ matrix, which to 2PN order is given by: 
\begin{equation}
\Psi = 
\left [ \begin{matrix}  
	  a_{01} 
	& 0 
	& {a_{21}}/{\theta_2} + ({a_{22}}/{3}) \left ( {\theta_2}/{\theta_1}
\right )^{2/3} 
	& 0 
	& {a_{41}}/{\theta_2^2} + {a_{42}}/\left ({3 \left ( \theta_1^2\theta_2
\right )^{1/3} } \right ) 
	- ({a_{43}}/{3}) \left ( {\theta_2}/{\theta_1} \right )^{4/3} \cr 
	  0
	& 0 
	& - {a_{21}\theta_1}/{\theta_2^2} + (2 {a_{22}}/{3}) \left (
{\theta_1}/{\theta_2} \right )^{1/3} 
	& a_{31} 
	& - {2a_{41} \theta_1}/{\theta_2^3} - ({a_{42}}/{3}) \left (
{\theta_1}/{\theta_2^4} \right )^{1/3}  
	+ ({4a_{43}}/{3}) \left ( {\theta_2}/{\theta_1} \right )^{1/3}
\end{matrix}
\right ].
\end{equation}

It is useful to note that the moment functional of polynomial 
functions $a(x)=\sum a_k x^k$ are given by
\begin{equation}
\mathcal{J} \left[ \sum_{k} a_{k} x^{k} \right] = \sum_{k} a_{k} J(7-3k),
\end{equation}
where $J(q)$ is the normalized moment given by $J(q)=I(q)/I(7).$
Using the definition of the metric in Eq.~(\ref{eq:gamma metric})
and projecting out the parameter $t_C$, we find 
\begin{equation}
\label{eq:metric}
g_{ml}  = \frac{1}{2}\sum_{k,j=0}^N \Psi_{mk} \Psi_{lj} 
\left  \{ J(17-k-j) - J(12-k) J(12-j) 
	- \frac { \left [ J(9-k) - J(4)J(12-k) \right ]
		  \left [ J(9-j) - J(4)J(12-j) \right ]} {\left [J(1) - J(4)^2 \right]}
\right \}.
\end{equation}

\section{Template bank based on SPA model}
\label{sec:template placement}

In this Section we will discuss the problem of constructing a bank of templates
based on the geometric formalism introduced earlier. To this end we shall use
the specific signal model of an inspiral binary discussed in
Sec.~\ref{sec:spa}. This model uses a restricted PN approximation in
which the PN amplitude corrections containing higher order harmonics
are neglected. The end result of the construction of the template bank is a set
of points in the parameter space of chirptimes or, equivalently, the masses.
Each point is associated with a template built from a specific signal model. 

In principle, there is nothing in the construction of the bank that forbids us
to use as our templates a model that is different from the one discussed in
Sec.~\ref{sec:spa}. Indeed, we shall show in a companion paper that although 
we have used a specific signal model in the construction of the bank, the 
same bank works for a wide variety of other signal models. In this paper we 
shall mainly focus on the SPA model, and only as an 
introduction to a more exhaustive study we shall present one case of a 
physical model based on the time domain Taylor model at 2PN, restricted 
to the BNS search.

We validate the performance of our template bank and quantify its 
{\it efficiency} (see Sec.~\ref{sec:efficiency} for a precise 
definition) using Monte Carlo simulations in which a number of signals, 
with their parameters chosen randomly, are 
generated and their best overlap with the bank of templates is 
computed. The algorithm described here is implemented in the \ac{LAL}
\cite{LAL}, and currently used by the LIGO Scientific Collaboration 
(LSC) to search for BNS, BBH and binary PBHs \cite{LIGOS2bns,LIGOS2bbh,LIGOS2PBH}.

\subsection{Effective dimensionality of the parameter space}
In order to construct the template bank we will start 
with the metric defined in Eq.~(\ref{norm}). As argued in
Sec.~\ref{sec:filtering} the extrinsic parameters
$t_C$ and $\varphi_C$ do not need to be searched with a template
bank. One can analytically maximize the overlap with respect
to these parameters locally at each point in the space of masses.
Thus, we project $\varphi_C$, $t_C$ onto the two-dimensional space
of masses in which the chirptimes ($\tau_0, \tau_3)$ defined by
Eq.~(\ref{eq:chirp parameters in masses}) serve as coordinates. Instead of
$(\tau_0,\tau_3)$ we can equivalently take any two of the $\tau_k$'s to be
independent parameters to characterize the signal. The choice ($\tau_0,
\tau_3$), is particularly attractive because their relationship to the masses is
analytically invertible \cite{Mohanty} [cf. Eqs.~(\ref{eq:masses in chirp parameters})
and (\ref{eq:chirp parameters in masses})]. In addition, as we will argue later,
the metric in these coordinates is a slowly varying function \cite{TanakaTagoshi}.  

\subsection{Lower frequency cutoff}\label{subsec:fl}
Before presenting an algorithm to place templates in the parameter
space let us discuss how one can choose the lower frequency cutoff $f_L,$
which essentially determines the size of the parameter space of chirptimes and
plays a crucial role in the computational resources required to process the data
through the template bank. 

The initial fiducial frequency $f_L$ defines the 
range of values of the chirptimes and is not itself a parameter to search for. 
However, it affects the length of the signals (therefore, the parameter 
space to be covered) and the SNR extracted.  For example, the 
Newtonian chirptime $\tau_0 \propto f_L^{-8/3};$ thus,
lower values of  $f_L$ give  longer templates with the immediate consequence 
of enhancing both the required storage space for templates and the 
computational cost to filter the data through the template bank. Furthermore, 
the number of cycles in a template increases as $N_{\rm cyc} \propto
({\cal M}f_L)^{-5/3},$ affecting both the overlap between different
templates and the number of templates in the bank.

What is really important, both for signal recovery 
and the number of templates required in the bank, is the 
effective number of cycles \cite{DIS2} which depends not only on 
the lower frequency cutoff but also on the noise PSD of the detector.
The reason for this is the following: while the signal power increases
with decreasing lower-cutoff as $f_L^{-7/3},$ the noise PSD $S_h(f)$
increases much faster below a certain frequency determined by the
various noise backgrounds. Thus, there will be negligible 
contribution to the SNR integral and noise moments $J(q)$ from 
frequencies below a certain lower cutoff. As a consequence, 
choosing the lower cutoff frequency to be smaller than a certain 
value has no advantage; on the contrary it has the undesirable effect 
of increasing the number of templates and the computational cost of the analysis.

The constant minimal match $MM$ has also to be set and we decided to chose a value $MM=95\%$.
This value is large enough so that Eq.~\ref{eq:overlap} is still valid and small enough so that the number 
of templates (Eq.~\ref{numTemp}) is not too large with respect to computational cost.

Table \ref{tab:banksize} summarizes the number of templates at a 
constant minimal match of $MM=95\%$ as a function of $f_L$. We can immediately
see that for the different PSDs considered, the number of templates evolves
significantly as $f_L$ is reduced but stabilizes after reaching a certain
value.  Our aim is to choose the lower frequency cutoff small enough so
that it has negligible effect on signal recovery.

\begin{table}
\caption{\label{tab:banksize} The number of templates required to search
for BNS versus the lower cutoff frequency $f_L$ at a minimal match 
$MM=95\%$. As the cutoff is lowered the number of templates initially 
increases sharply but stabilizes after reaching a certain value 
depending on the noise characteristics of the detector in question. The
last row gives the lower cutoff frequency that was chosen in our
simulations by demanding that the loss in SNR due to this choice is
not more than 1\%. Note that the number of templates converges at smaller
values of $f_L,$ but we don't gain in SNR by choosing $f_L$ to be
that at which the number of templates converge.} 
\begin{ruledtabular}
\begin{tabular}{ddddd}
\text{$f_L$\rm(Hz)} & \text{GEO\,600} & \text{LIGO-I} & \text{Advanced LIGO} &
\text{Virgo} \\ \hline
10 & 6085 & 4382 & 18692 & 30650\\ 
15 & 6099 & 4412 & 12782 & 22068\\ 
20 & 6113 & 4449 & 9425 & 15992\\ 
25 & 6129 & 4408 & 7086 & 12027\\
30 & 6145 & 4312 & 5469 & 9140 \\
35 & 6002 & 3783 & 4312 & 6958 \\
40 & 4914 & 3121 & 3527 & 5504\\
45 & 3768 & 2577 & 2885 & 4358\\
50 & 3017 & 2271 & 2289 & 3458 \\ \hline
\text{Chosen $f_L$\rm(Hz)} & 40 & 40&  20 & 20 
\end{tabular} 
\end{ruledtabular}
\end{table}

Choosing the lower frequency cutoff too high leads to a loss in 
the SNR because, as evident from Eq.~(\ref{eq:SPA}), there 
is greater power in the signal at lower frequencies, with the 
power spectrum of the signal falling with frequency as $f^{-7/3}.$ However, 
the increase in signal power at lower frequencies does  not help beyond a 
certain point because the SNR integrand is weighted down by the noise 
PSD of the instrument. Current generation of interferometers are dominated 
 at frequencies below about $30$-$40$ Hz by seismic noise. For instance, in the
case of the Virgo interferometer the noise PSD raises as $f^{-4.8}$ at
frequencies below about $30\rm Hz$ thereby dictating that setting
$f_L \sim 30 \rm Hz$ might be a reasonable choice. 

Essentially, there are two opposing choices for $f_L$ and one must 
make an arbitrary but optimal choice. In this paper, the lower frequency cutoff
is chosen as large as possible
with the constraint that no more than 1\% of the 
overlap is lost as a result of choosing $f_L$ to be different from 0.
The choice of $f_L$ would then depend, in principle, on the masses
of the stars in the binary.  This is because the templates are shutoff 
when the binary reaches the innermost circular orbit,
making the bandwidth available to integrate the signal smaller, 
and forcing to choose a lower cutoff that is smaller for 
higher masses. However, again because of the steep increase in the
seismic noise below a certain frequency, signals from binaries 
of total mass greater than $\sim 100 M_\odot$ will not be
visible in the first generation instruments. 

\begin{figure}[tbh]
\centering
{\includegraphics[width=0.45\textwidth,angle=0]
{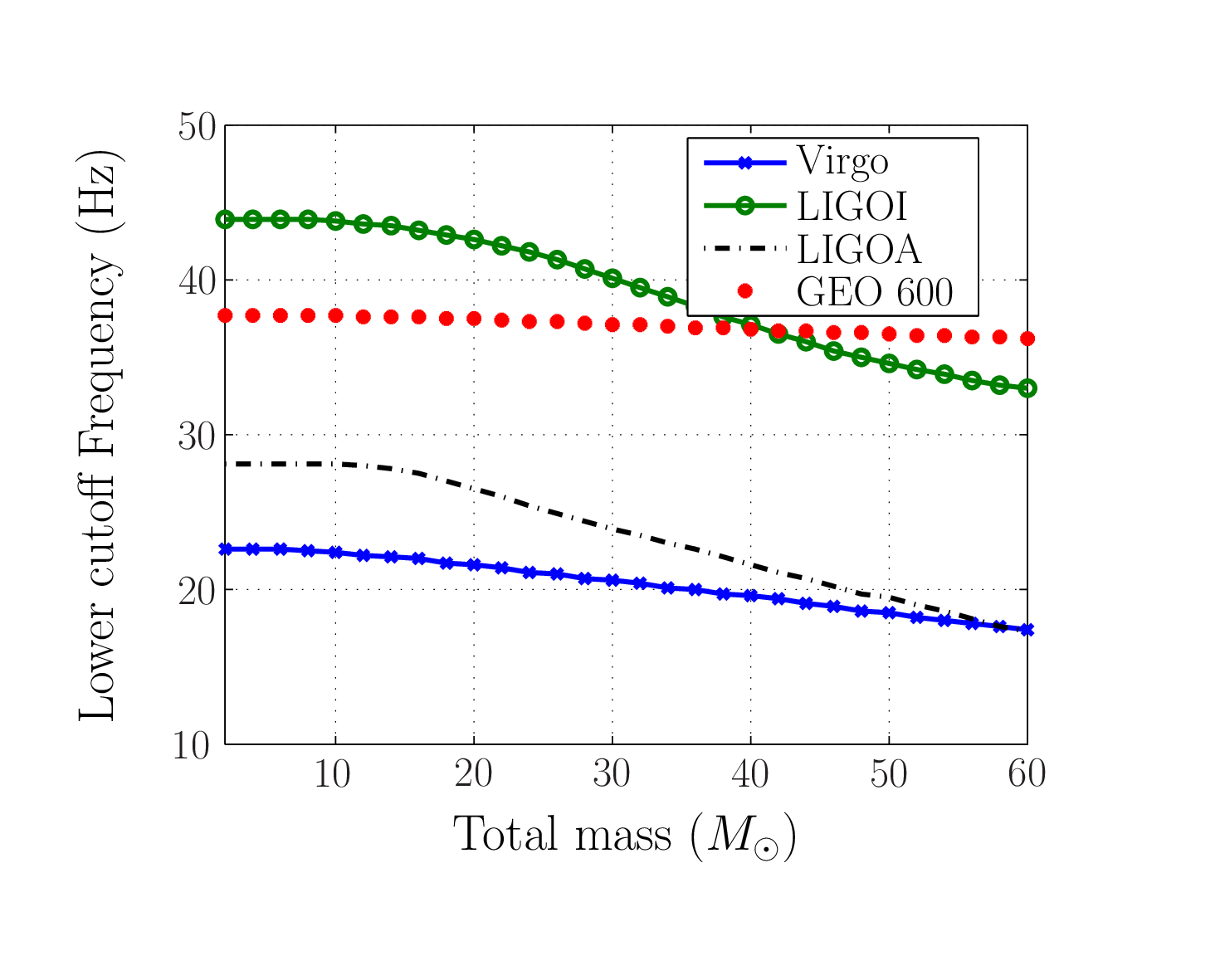}
\includegraphics[width=0.45\textwidth,angle=0]
{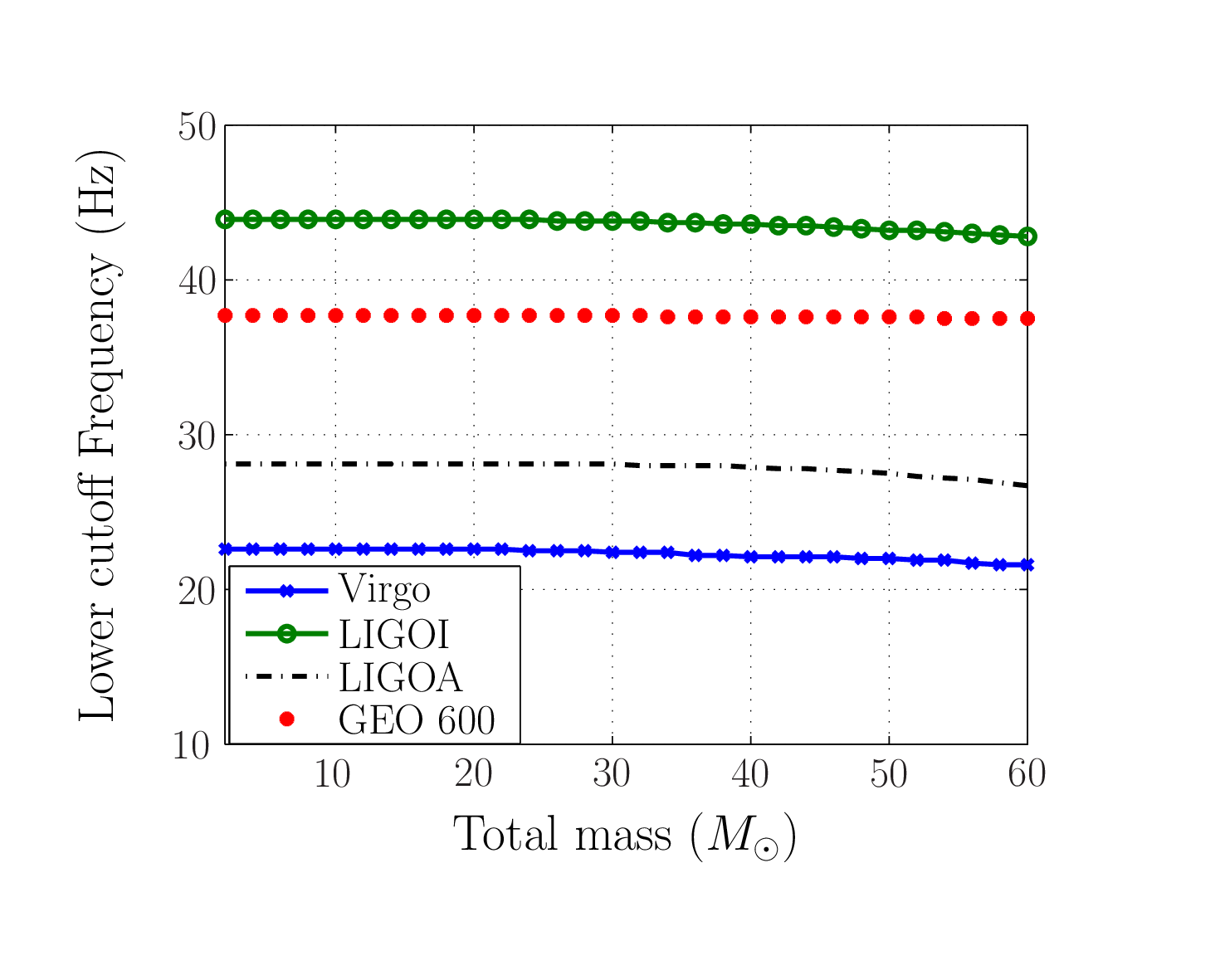}}
\caption{The four curves in each plot show the lower frequency 
cutoff that corresponds to a loss of SNR of 1\%, respectecly for the 
GEO\,600, LIGO-I, Advanced LIGO and Virgo, using the ending frequency of 
the binary to be the last stable orbit (left panel) and the light ring
(right panel).}
\label{fig:flower}
\end{figure}

In Fig.~\ref{fig:flower} we plot the highest value of lower 
cutoff as a function of the total mass of the binary by requiring
that the loss in the SNR is no more than 1\%. The different
curves in the plot correspond to the noise PSD expected in various 
ground-based interferometers. There is no need to make these plots
for systems with different mass ratios since, at the level of
approximation we are using, the cutoff depends only on the total
mass and not on the mass ratio of the system. The left panel plots the
lower cutoff frequency, which gives a loss in SNR of 1\%, as 
a function of the total mass when the ending frequency 
of the system is set at the last stable orbit\footnote{The
last stable orbit frequency is  
taken to be the one given for a test mass in Schwarzschild geometry: 
$f_{\rm LSO}=1/(6^{3/2} \pi M)$} (LSO). The right panel shows
the same result when the upper frequency is set at the light ring 
(EOB models at 2PN). Our choice of lower frequency cutoff is 
summarized in Table~\ref{tab:banksize}.

There is subtlety \cite{OwenPrivate} in the choice of the
lower-frequency cutoff.  The optimal SNR is only sensitive to the 
lower-frequency cutoff but not the parameters of the template and the 
signal because they are both assumed to be the same. When filtering the
data through a template bank, however, we have no optimal templates but for
a set of measure zero signals. Consequently, what is important 
is not the behaviour of the SNR (which only depends on
the moment $J[7]$) but that of the metric (which is sensitive to the
difference in the parameter values of neighbouring templates). The
metric is highly sensitive to the moment $J[17],$ which converges
rather slowly as a function of the lower-cutoff.  We have chosen
the easier approach described above in choosing the lower-cutoff,
as that is less sensitive to the parameters of the system. 
The real justification comes from the fact that our choice of the 
lower-cutoff does not seem to have severely affected the 
efficiency of the template bank as we shall see later in this paper.

\subsection{Template placement algorithm}
The templates are chosen in the parameter space such
 it is guaranteed that for every
signal in the region of interest there will be at least one template 
in the bank with an overlap larger than or equal to the minimal
match.  In addition to the templates within the parameter space 
of the search a layer of templates might also be required just 
outside the boundary of the parameter space to assure the chosen 
minimal match for all signals within the parameter space. This
 happens because some of the templates that lie outside the 
boundary have coverage within the parameter space; without them
signals with their parameters near the boundary might not have the
desired minimal match \cite{OwenPrivate}.

The template placement algorithm consists of two steps: the first 
step deals with placing templates along a chosen curve in the
parameter space, specifically along the equal-mass curve,
while the second step deals with the placement of
templates within and outside the boundary of the parameter 
space.

The preference of chirptimes (or chirp parameters) over masses 
as coordinates on the signal manifold is dictated by the fact 
that these variables are almost Cartesian. This is clearly 
seen when considering the 1PN model for 
which the metric $g_{\mu\nu}\equiv \left ( h_\mu,\, h_\nu \right )$ 
is a constant, i.e.\ independent of the chirptimes, while the
same metric expressed in $(M, \eta)$ or $(m_1,m_2)$ coordinates
is not a constant. Thus,
the signal manifold at 1PN order is not only flat, the
chirptimes are Cartesian-like coordinates. When we go to higher
PN orders, the multi-dimensional signal manifold, in which all chirptimes
are considered to be independent of one another, is again a 
flat manifold but the physical chirp manifold (i.e.\ the 
manifold formed by signals expected from the inspiral of 
black hole binaries) is only a two-dimensional manifold that
is obtained by imposing the constraints $\tau_3=\tau_3(\tau_0, \tau_1)$,
$\tau_4=\tau_4(\tau_0, \tau_1)$, and so on. This physical manifold,
could, in principle, be a curved manifold but the curvature is 
not likely to be large since the absolute value and range of
higher order chirptimes is small compared to the most dominant
chirptimes. Hence the constraints are unlikely to result 
in a large curvature.  We shall, therefore, assume 
that the metric is essentially constant  
in the local vicinity (on the scale defined by mismatch) of every point 
on the manifold.
\begin{figure}[tbh]
\centering
\includegraphics[width=0.45\textwidth,angle=0]
{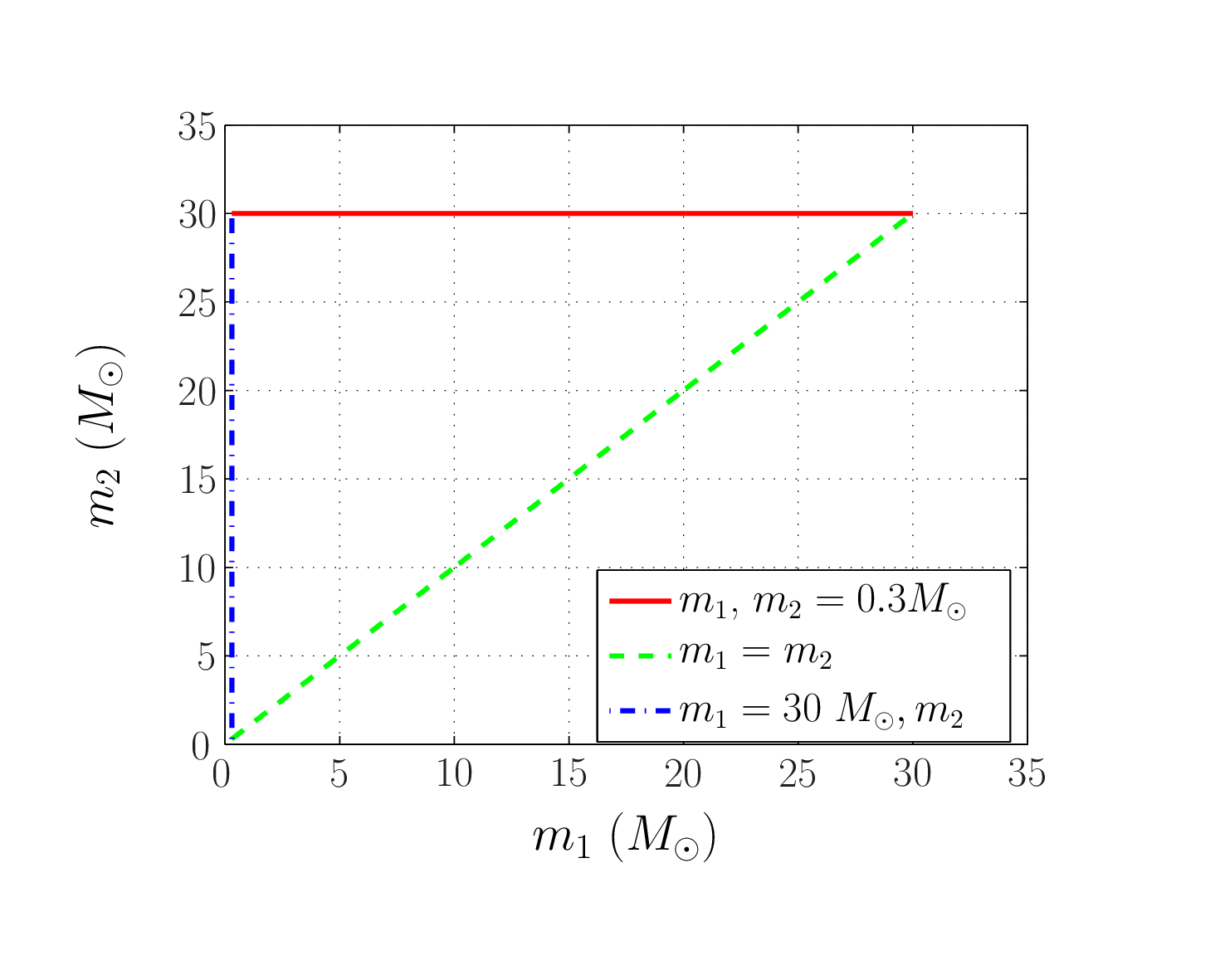} 
\includegraphics[width=0.45\textwidth,angle=0]
{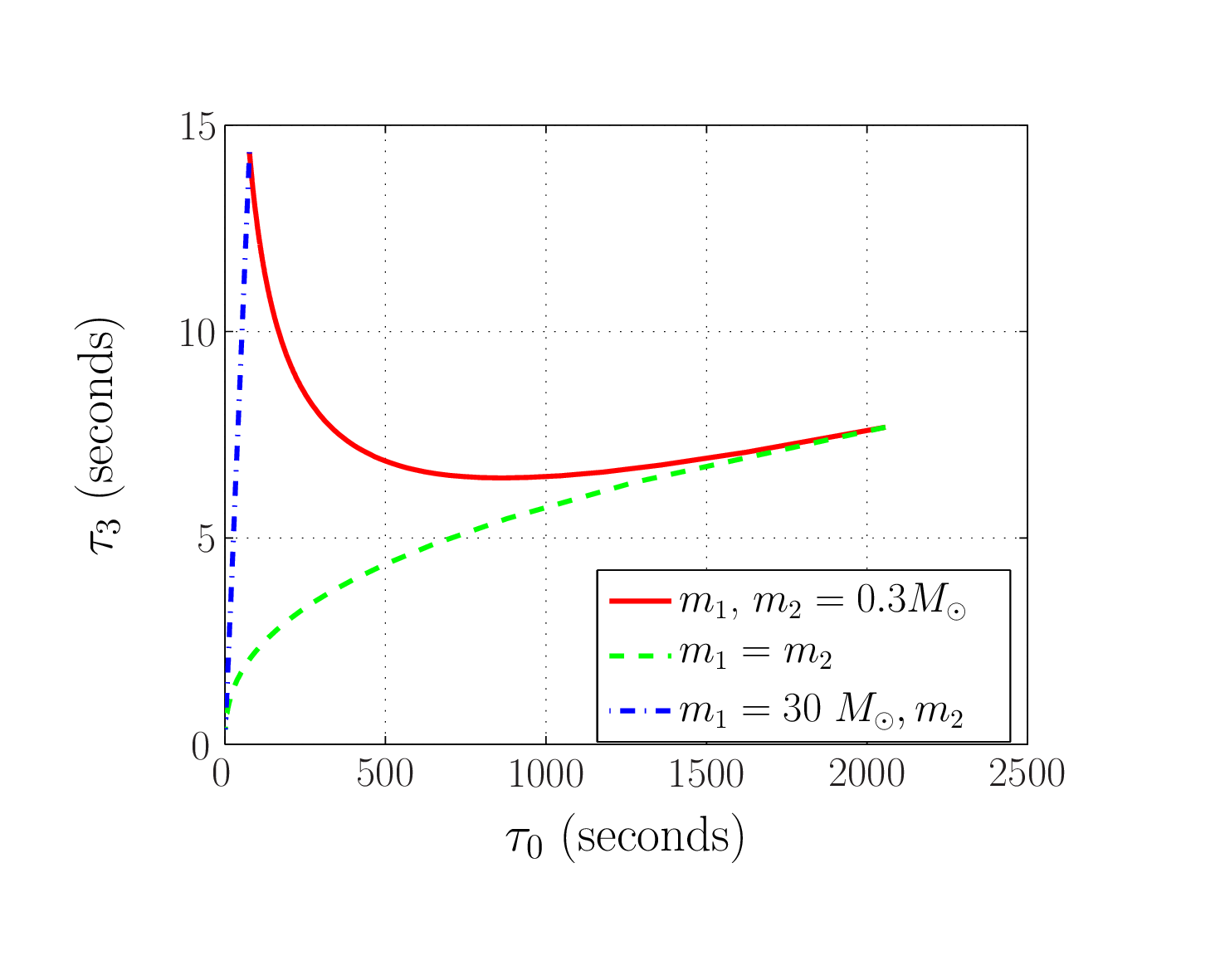}
\caption{The left panel shows an example of the range of masses 
covered by the template bank while the right panel shows 
the corresponding region in the space of chirptimes 
$\tau_0$-$\tau_3.$ The shape of
the detector's PSD does not enter in the definition of the chirptimes but the
lower frequency cutoff does
and is taken to be $f_L=20$ Hz. The individual
mass range is chosen to be [0.3,~30]$M_\odot$ including the BNS, BBH and PBHs
binary systems.}

\label{fig:t0t3}
\end{figure}

In this work we shall specifically choose $(\tau_0,\, \tau_3)$
to be the coordinate system. As we also mentioned above, the masses 
are analytically invertible in terms of this pair  of chirptimes
[cf.\ Eq.~(\ref{eq:masses in chirp parameters})]. The range
of masses accessible to various interferometers is shown 
in Fig.~\ref{fig:t0t3} together with the corresponding
space of chirptimes. Note that since $\eta \le 1/4$ not all
of the $M$-$\eta$ or $\tau_0$-$\tau_3$ space
is physically meaningful. In the $M$-$\eta$ space the region
above $\eta=1/4$ is forbidden while in the space
of $\tau_0$-$\tau_3$ the region below the curve marked
$m_1=m_2$ is forbidden\footnote{Owen \cite{OwenPrivate}
has suggested that it might be worthwhile to express the
signal entirely in terms of $M$ and $\eta$ (or entirely 
in terms of $\tau_0$ and $\tau_3$) and explore the
forbidden region. This is a way of expanding our net to
catch the inspiral signals whose late time phase evolution
is not accurately described by the PN expansion.
This suggestion was independently explored and extended by
Buonanno, Chen and Vallisneri \cite{BCV1}}.
As we do not wish to place templates in this forbidden region
(although this is, in principle, perfectly possible) the loss
in SNR will be more than the minimal match if we do not place
templates along the equal-mass curve. Imagine starting 
from the boundary on the left of the space of chirptimes and
placing templates along the $\tau_3=\rm const.$ line. Eventually,
the algorithm would take us beyond the boundary on the right
and a proper coverage of the parameter space would require
us to place a template in the forbidden region. Placing this
last template on the equal-mass curve is not a solution as
this could create a ``hole'' in the parameter space. For
signals in the ``hole'' the overlap obtained will be less
than the minimal match.

Thus, our template placement algorithm consists of two steps: 
After computing the minimum and maximum chirp-times corresponding to the
search space, namely $(\tau_0^{\rm min}, \tau_0^{\rm max})$
and $(\tau_3^{\rm min}, \tau_3^{\rm max}),$ the algorithm
chooses a lattice of templates along the equal mass curve,
lays a grid of templates in the rectangular region defined by
the minimum and maximum values of the chirptimes and rejects
the lattice point if the point itself lies outside the parameter 
space {\bf and} none of the vertices of the 
rectangle inscribed within ambiguity ellipse lie 
within the parameter space.

\subsubsection{Templates along the $\eta=1/4$-curve}
\begin{figure}[htb]
\begin{center}
\includegraphics[angle=0,width=0.45\textwidth]
{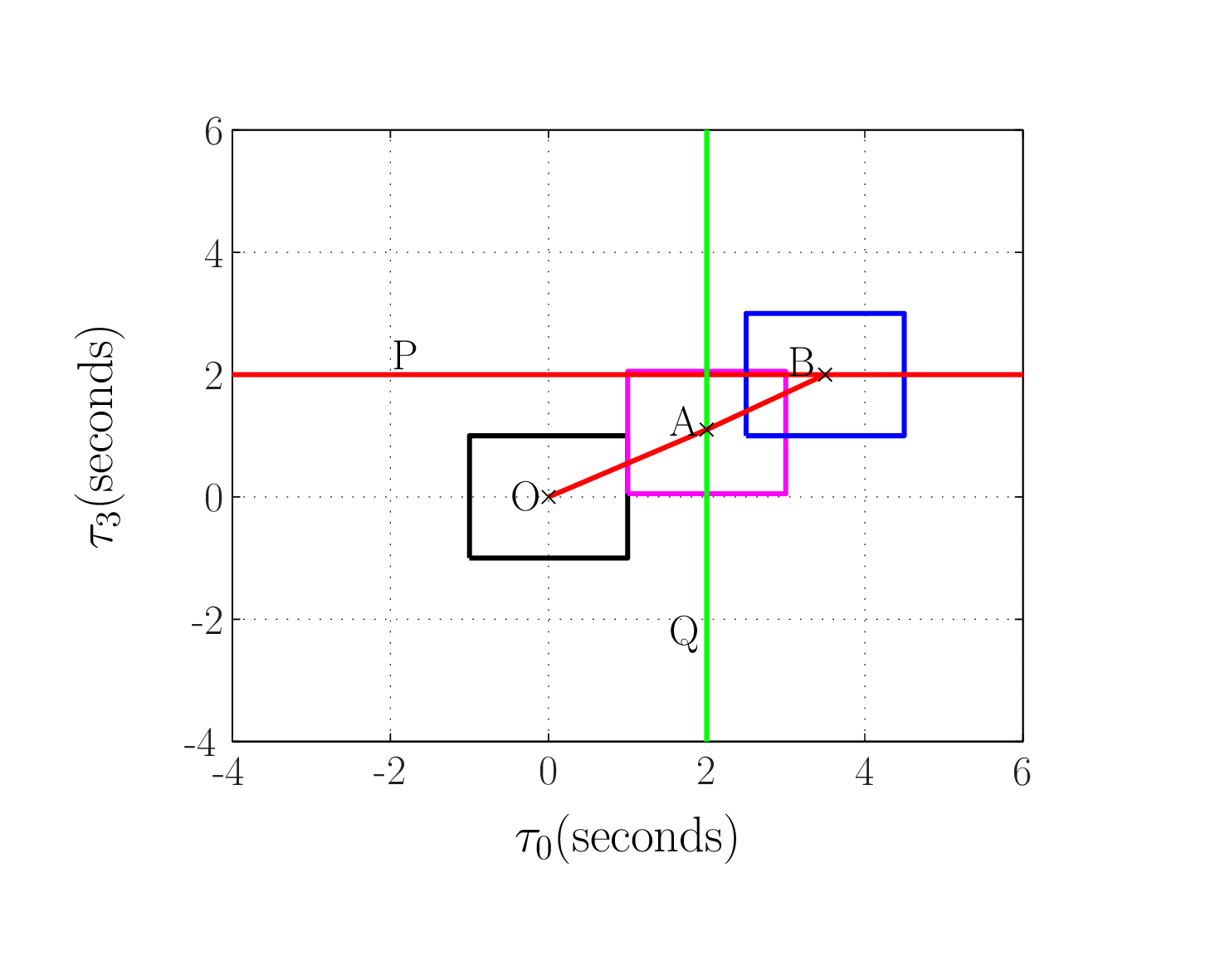}
\end{center}
\caption{Algorithm sketching the placement of templates along the
$\eta=1/4$ curve. The placement uses the rectangle inscribed in the ambiguity
 ellipse around each point of the parameter space considered. }
\label{fig:equal mass algo}
\end{figure}

In the first stage, templates are built along the equal
mass (that is, $\eta=1/4$) curve starting from the minimum value
of the Newtonian chirp-time and stopping at its maximum value. 
The algorithm is illustrated in Fig.~\ref{fig:equal mass algo}:
given the $n$-th template at $O$ with parameters $(\tau_0^{n},\tau_3^{n}),$
and the distance $(\delta\tau_0^{n},\delta\tau_3^{n})$ 
between templates in our preferred coordinates,
consider lines $\tau_0 = \tau_0^{n} + \delta\tau_0^{n}$ 
($QA$ in Fig.~\ref{fig:equal mass algo}) and
$\tau_3 = \tau_3^{n} + \delta\tau_3^{n}$ 
($PB$ in Fig.~\ref{fig:equal mass algo}).
The template next to 
$(\tau_0^{n},\tau_3^{n}),$ on the equal mass curve, must lie
either along $PB$ or along $QA$ (cf. Fig.~\ref{fig:equal mass algo}) in order
that all the signals that may lie on $OAB$ 
are spanned by at least one of the two templates.  Clearly, if we were
to place the $(n+1)$-th template at $B$ there will be a gap and
some of the signals will not have the required minimal match; 
placing it at $A$ meets our requirement. 

Note, however, that there is no guarantee that this will always work; 
it would indeed fail if the curve along which templates are being laid is a 
rapidly varying function. However, there is no danger of this happening
in the case of the $\eta=1/4$ curve. Indeed, for increments $\delta\tau_0$
equal to the distance between templates  expected at a minimal match 
of 0.95 we can assume the curve to be a straight line: 
$\tau_3 \approx \tau_3^0 + \dot{\tau}_3^0 \delta \tau_0,$ where $\dot{\tau}_3$
is the derivative with respect to $\tau_0$ along the $\eta=1/4$ curve.
Therefore, the above algorithm will never fail in our case.

To locate the $(n+1)$-th template we compute the following pairs 
of chirptimes:
\begin{eqnarray}
\tau_0^{n+1} = \tau_0^{n} + \delta\tau_0^{n}, \ \ 
\tau_3^{n+1} =  4A_3 \left ( \frac{\tau_0^{n+1}}{4A_0} \right )^{2/5}
\nonumber \\
\tau_3^{n+1} = \tau_3^{n} + \delta\tau_3^{n}, \ \ 
\tau_0^{n+1} =  4A_0 \left ( \frac{\tau_3^{n+1}}{4A_3} \right )^{5/2},
\end{eqnarray}
where 
\begin{equation}
A_0=\frac{5}{256 (\pi f_L)^{8/3}}, \ \ A_3=\frac{\pi}{8 (\pi f_L)^{5/3}}.
\end{equation}
Of the two pairs, the required pair is the one that is closer to the 
starting point $(\tau_0^{n},\tau_3^{n}).$ 

\subsubsection{Templates in the rest of the parameter space}
In the second stage, the algorithm begins again at the point 
$(\tau_0^{\rm min}, \tau_3^{\rm min}),$ 
with the corresponding distance between templates
$(\delta\tau_0^{\rm min}, \delta\tau_3^{\rm min}),$ and chooses 
a rectangular lattice of templates in the region defined by 
$(\tau_0^{\rm min}, \tau_3^{\rm min})$ 
$(\tau_0^{\rm max}, \tau_3^{\rm min})$ 
$(\tau_0^{\rm max}, \tau_3^{\rm max})$  and
$(\tau_0^{\rm min}, \tau_3^{\rm max})$. 
A template is accepted either if its coordinate is within the
parameter space of search or if its span, for the given minimal
match, has an overlap with the parameter space. It is this latter
requirement that allows for a layer of templates just outside the
boundary of the parameter space. These templates will
have overlaps larger than or equal to $MM$ for some signals 
within the parameter space of interest; leaving them out would
cause `holes' in the template bank where the match achieved
by the template bank will be smaller than the minimal match.

\begin{figure}[h]
\begin{center}
\includegraphics[angle=0,width=0.5\textwidth]
{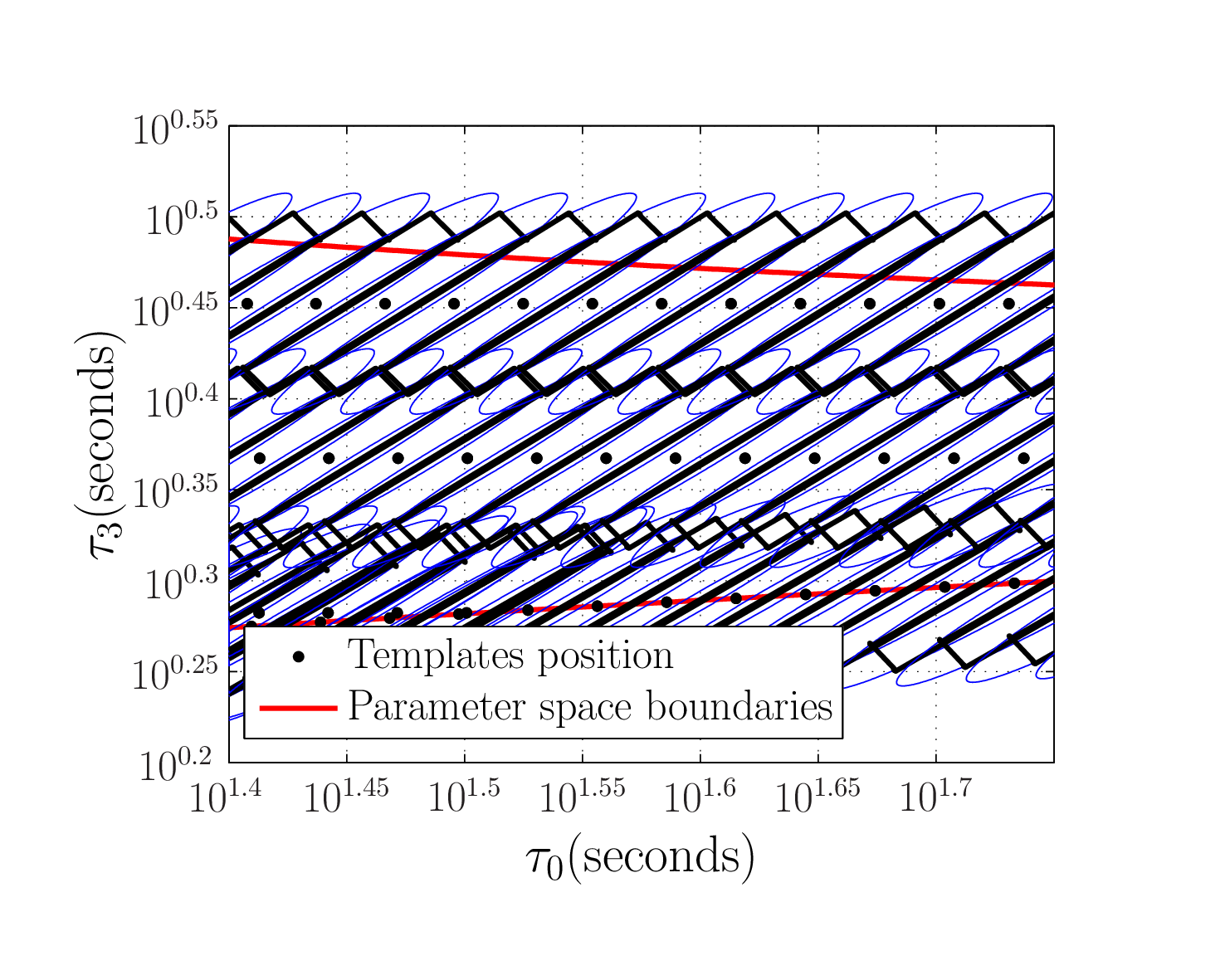}
\caption{Algorithm sketching the construction of a rectangular lattice of templates. Using the placement algorithm as shown in Fig.~\ref{fig:equal mass algo}, the parameter space can be spanned with ellipses using their inscribed rectangles.}
\label{fig:coarse}
\end{center}
\end{figure}

The implementation of the algorithm along the equal mass curve and
in a rectangular lattice in the rest of the parameter space is 
plotted in Fig.~\ref{fig:coarse} where the chosen templates are 
represented as points and the ellipses around the points. The ellipses define the regions where the overlap is greater than or equal to 
the minimal match. A sketch of the implementation of the
algorithm is given in Appendix \ref{sec:app} for both the placement 
along the equal mass curve and the rest of the parameter space.

In summary, the following parameters are required to compute the grid 
in the parameter space:
\begin{itemize}
\item {\it minimum mass} of the component stars,
\item {\it maximum total mass} of the binary system,
\item {\it minimal match} which defines the distance between 
the nearby templates in the parameter space, 
\item {\it noise power spectral density} $S_h(f)$ of the detector in question,
\item {\it lower} and {\it upper frequency cutoffs} 
used in computing the moments of the noise PSD.
\end{itemize}

\section{Validation of the template bank}\label{sec:efficiency}

We have carried out Monte-Carlo simulations to study the efficiency of the 
template bank described in the previous Section. By efficiency we 
mean the ability of the bank to capture
signals with the loss in SNR no more than the one defined by $MM$, the
minimal match. The Monte-Carlo simulations consist in generating 
a number $N_s$ of random {\it normalised} signals  $\hat{s}$ and
finding the fitting factor $FF$ between the signal and the bank of templates.
We use $\hat{s}$ to denote the signal in order to distinguish it from the
template $\hat{h}$ used in the bank. The models used to generate $\hat{s}$ and
$\hat{h}$ are based on the SPA model at 2PN described in the Section~\ref{sec:spa}. 
We randomize the parameters ($m_1,
m_2$) in the range covered by the template bank as well as the  initial
phase $\varphi_C$. Another issue is the randomization of the mass parameters,
for which there are several choices: one choice could be to have a
uniform distribution of component masses. However, this choice will not 
probe our template bank properly; indeed, the corresponding 
distribution of the total mass will be under-populated in the small 
or the large total mass range.
In fact, in order to properly test a template bank, we should
inject signals using uniform distribution of the parameters used to design the
bank. Namely the set $(\tau_0, \tau_3)$. But again such a choice will
over-populate the lower-end of the physical masses. 
We found it more convenient to inject
signals with a uniform distribution in the total mass of the binary, although
this choice is not necessarily astrophysical. Moreover, the total mass
determines the frequency at which the PN waveform terminates and it would,
therefore, be helpful in choosing our injections based on a uniform
distribution in the total mass. Such a distribution would emphasize any problems with the efficiency of the template bank
associated with waveforms that terminate in the sensitive bandwidth of the
detector and/or have short duration.

\subsection{Figures of merit}
From the results of the Monte-Carlo simulations we construct several 
figures of merit (FOMs) that are relevant in validating the template bank.
First, we define the efficiency of the template bank by a vector $\mathcal{E}$
defined by 
\begin{equation}
\mathcal{E}_{N_s}\left(
\mathcal{\chi}_{s},
 \mathcal{\chi}_{h}\right)
= \left\{
FF\left(\hat{s}_i(\vartheta^s), h(\vartheta^h)  
\right)
\right\}_{i=1..N_s}
\end{equation}
where $\vartheta^{s}$ and $\vartheta^{h}$ are vectors corresponding to
the parameters of the injected signals and the templates, $\chi_s$
and $\chi_h$ are the models used in the generation
of the signal and template, respectively.  In all the simulations, we set 
$\vartheta^{s} = \{m_1, m_2, \varphi_C\},$ but as
mentioned 
earlier we can analytically maximise over the orbital phase $\varphi_C$ and, 
therefore, $\vartheta^{h} = \{m_1, m_2\}$. 
Moreover, we fix $\chi_s = \chi_h = \rm{SPA}$ model. 

From the efficiency vector $\mathcal{E}$ and the injection parameter vector
$\vartheta^s$, we can derive several FOMs
\begin{enumerate}
\item $\mathcal{E}$ versus the total mass. This is an important FOM
which shows (1)~if any injection has a value of $FF$ less than the specified $MM$
and (2)~if so,  for which range of total mass it occurs (see
Fig.~\ref{fig:efficiencyMhisto}).
\item A drawback of the first FOM is that it hides the information
about the mass ratio $\eta$ and its relationship to the efficiency.
Although not important in the BNS case, where the minimum value of
$\eta$ is equal to 0.1875, it can be interesting in other cases to look at the
dependency of the efficiency on the $\eta$ parameter. 
For instance, in the case of a binary consisting of a black hole 
and a neutron star (BH-NS) or BBH where $\eta$ can be significantly
different from its maximum value of 1/4.
\item A third interesting FOM is the cumulative distribution of $\mathcal{E}$
as in Fig.~\ref{fig:cumFF}. This is interesting since we can see how quickly
the distribution drops when  $FF$ is close to the minimal match. If it drops
sharply then it is a good sign telling us that the bank has the desired properties.
However, a smooth drop is a warning of the coarseness of the bank or the 
presence of holes in it. Indeed the distribution is expected to be quadratic.
\end{enumerate}

Finally, let us introduce a quantity called {\it safeness} defined as 
\begin{equation}
\mathcal{S}\left(
\mathcal{\chi}_{s},
\mathcal{\chi}_{h}\right) = \min \mathcal{E}(\chi_s, \chi_h).
\end{equation}
Since we can perform only a finite number of simulations, we have to take into
account the quantity $N_s$:
\begin{equation}
\mathcal{S}_{N_s}\left(
\mathcal{\chi}_{s},
\mathcal{\chi}_{h}\right) = \min_{N_s} \mathcal{E}_{N_s}(\chi_s, \chi_h).
\end{equation}
This is the minimal $FF$ found among the $N_s$ injections. Higher the $N_s$ 
the more confident we are with the value of the safeness. For instance, if
$N_s$ is set to 1\,000, we might miss some holes in the template
placement that might appear with a higher number of injections. If this quantity is less than the $MM$ then the bank
is under-efficient with respect to the minimal match chosen. Conversely, if
the $\mathcal{S} \gsim  MM,$ then the bank is said to be
efficient. The safeness is also dependant on the mass range considered. For
instance, a bank can be efficient in a subset $S$ of the mass range 
considered and under-efficient in its complement $S^C$. 
Any mass range in which the template bank is under-efficient 
should be considered separately and investigated more carefully. 

\subsection{Simulation parameters}
The parameter of each simulations are 
\begin{itemize}
\item  The mass range as defined by the minimum and maximum 
masses of the individual components of the binary. Although 
the parameter space as a whole need not be split into different
categories, we found it convenient to study four different cases: 
BNS, BBH, BH-NS and binary PBHs, each having different number of 
templates and efficiencies. Our placement algorithm has been
independently verified in the case of binary PBHs by
D.\ Brown \cite{LIGOS2PBH,BrownThesis} and found to be
efficient.  The mass ranges used in the different cases are
\begin{itemize}
        \item Binary PBHs: [0.3-1]$M_\odot$. 
        In principle, we can choose the lower
        limit to be a value smaller than 0.3~$M_\odot,$ but in practice 
        it implies a significant increase in the number of 
        templates than we can handle.  For example, in the case of LIGO-I,
        decreasing the lower limit to 0.2~$M_\odot$ and 0.1~$M_\odot$
        leads to roughly 130\,000 and 950\,000 templates, respectively.
	\item BNS : [1-3]$M_\odot$. 
	\item BBH : [3-30]$M_\odot$. We choose the upper limit to be 
        30 $M_\odot.$  The upper limit can be extended 
        depending on $f_L;$ in the case of advanced LIGO 
        or Virgo, we can expect to go up to 50 $M_\odot$. However, in
	our simulations we have chosen the same value for all the detectors.
	\item BH-NS [1-30] $M_\odot$ : we fix one of the component masses 
         to be in the range [1-3] $M_\odot$ and the other to be in the
         range [3-30] $M_\odot$. 
\end{itemize}
\item Design sensitivity curves are GEO\,600, LIGO-I, advanced LIGO and VIRGO  
as in Ref.~\cite{DIS2} have been used in our simulations.
Since PSDs used are design sensitivity curves,
the number of templates in each simulation remains the same
as opposed
to a Monte Carlo simulation involving real data (where the noise PSD
would change from one data segment to the next hence changing the number
of templates and their location in the parameter space).

\item The minimal match $MM$ is set at 95\% in all simulations. $MM$
parameter directly affects the number of templates in the bank
[cf.\ Eq.~(\ref{numTemp}].  Table \ref{tab:banksize2} summarizes 
the number of templates in the bank for 
each detector and search.

\item For each combination of design sensitivity curve and type of search
(BBH, BNS, \dots)  we carry out 
$N_s =10\,000$ injections.  

\item The sampling frequency is fixed at 4 kHz in all cases and, therefore,
the templates and signals are shut-off either at the last stable orbit
or the Nyquist frequency, whichever is smaller. 

\end{itemize}

\begin{table}[t]
\caption{The size of the template bank for the different searches and detectors.
The BH-NS case includes here the BBH and BNS. See
Fig.~\ref{fig:parameterspace} for a graphical representation.}
\label{tab:banksize2}
\begin{ruledtabular}
\begin{tabular}{ccccc}
Bank size versus detector & GEO\,600 & LIGO-I & Advanced LIGO& Virgo\\\hline
BBH&1219&774&2237&4413\\
BH-NS&9373&9969&67498&74330\\
BNS&5317&3452&9742&17763\\
Binary PBHs&56145&39117&115820&212864
\end{tabular}
\end{ruledtabular}
\end{table}

\subsection{Results and discussion}
Although the simulations we performed were split into 4 sub-categories,
there is nothing which forbids us to cover the whole parameter space in a single
category. However, in practice this has two major implications. First, for the
full range of masses there are far too many templates.
For instance, if we allow the template bank to cover the range defined by
both binary PBHs and BBH then the number of templates become very large
(160,000 in LIGO-I design sensitivity curve). Second,
splitting the template bank into sub-banks allows us to cover more extensively 
each specific binary and to tune the parameters used in each search. 
For instance, we can decrease the sampling
frequency of the BBH search from 4 kHz to 2 kHz without any loss in the SNR
because the signals in this region have an ending frequency below 1kHz. 

\begin{figure}[tbh]
\centering
{\includegraphics[width=0.45\textwidth]
{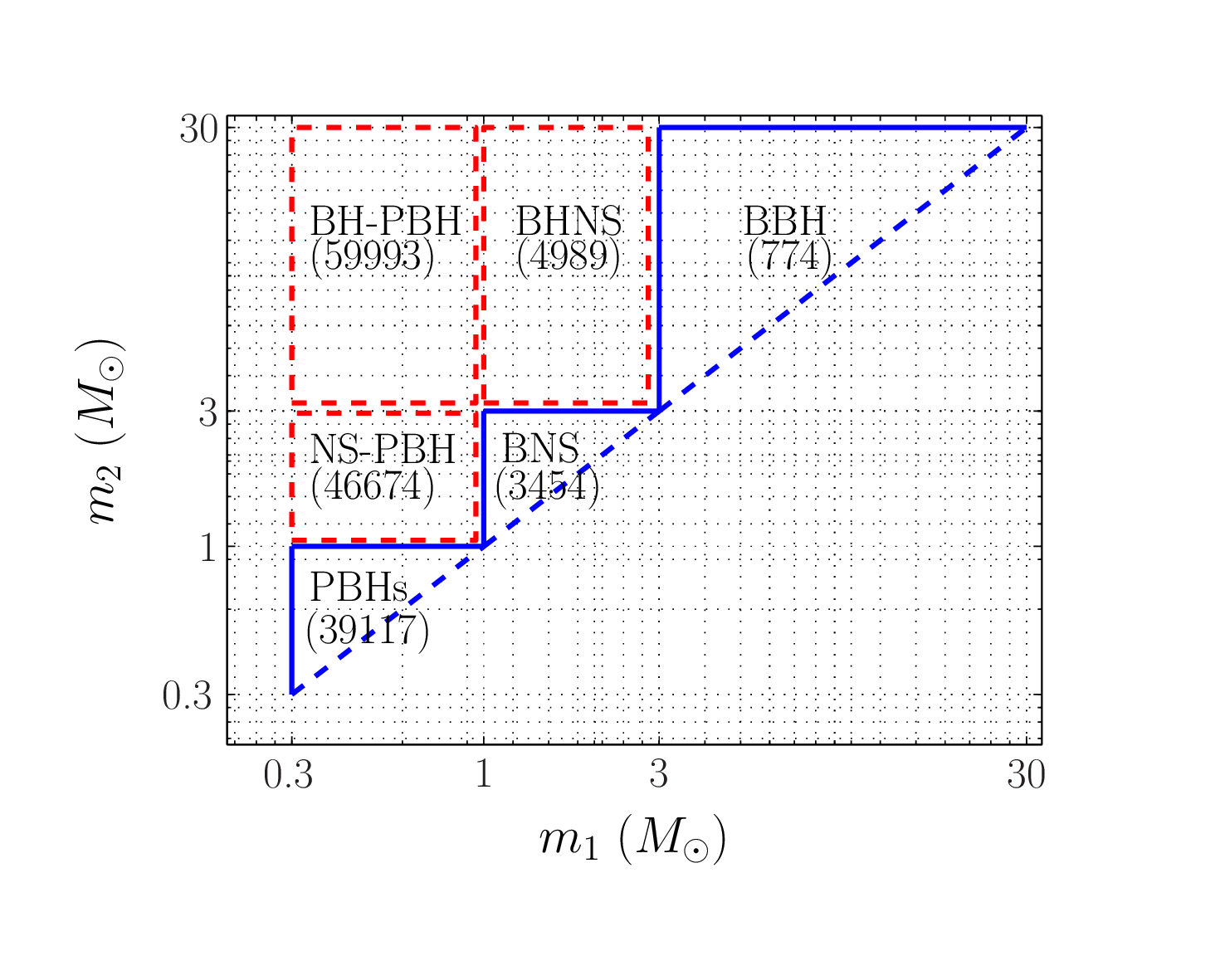}}
{\includegraphics[width=0.45\textwidth]
{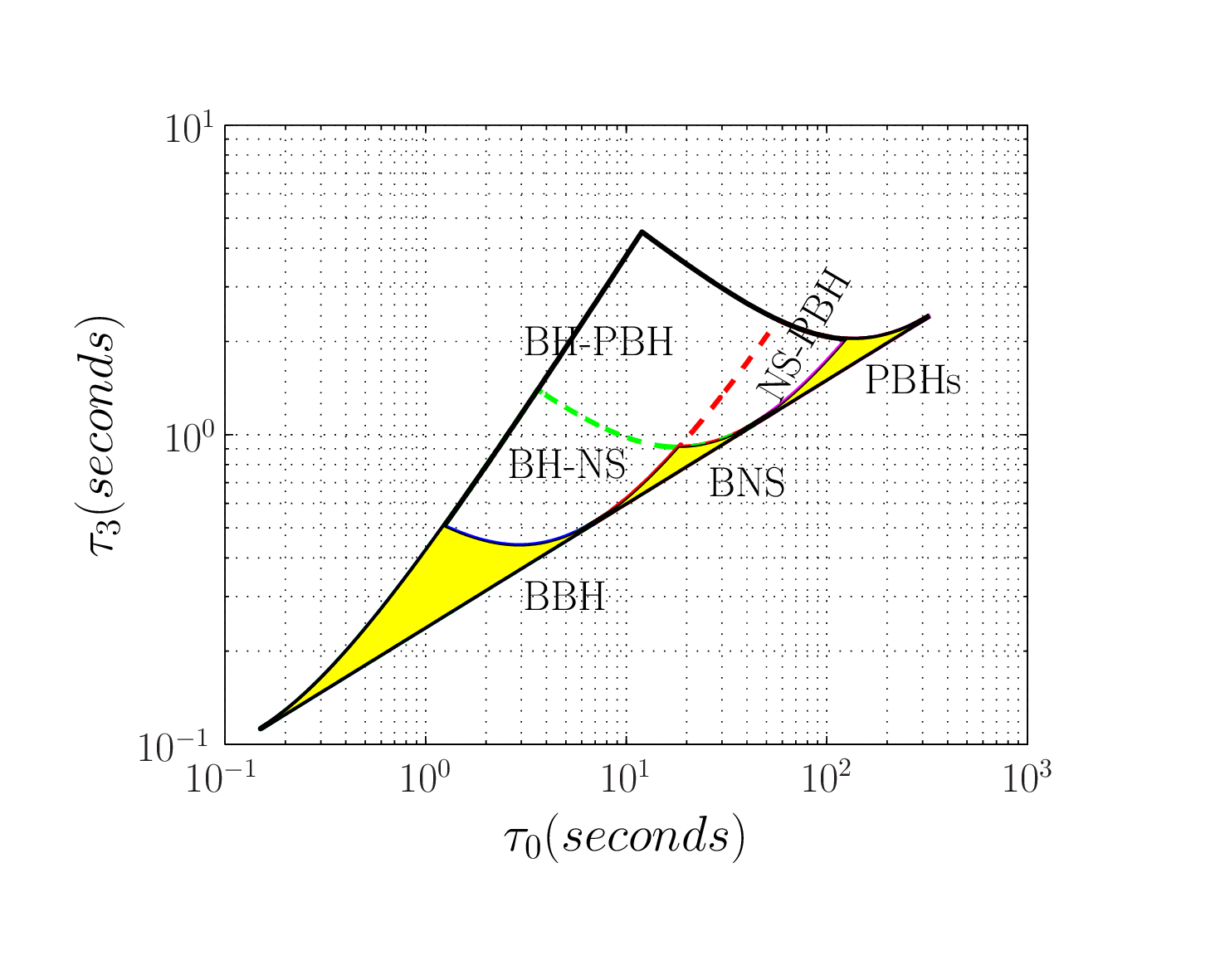}}
\caption{
The parameter space of the search can be split into sub-spaces corresponding to
astrophysically interesting binary sources. The left panel shows the
parameter space of masses of the component stars for various
astrophysically interesting systems while the right panel shows the
same in the space of chirptimes.
The properties of the compact object and the number of templates needed
to cover these sub-spaces are quite different. The solid \textbf{triangles} on the left plot
 show sub-spaces where both the objects are of the same type and in the same mass range. 
The dashed squares show binary systems in which the two objects belong to different
ranges of mass and are possibly of different type.  While binary PBHs, 
BNS, BBH and BH-NS are all astrophysically possible, binaries where
one of the components is a PBH and the other is a neutron star or a 
black hole probably don't exist and need not be searched in the first instance.
Note that in the space of chirptimes, where the area is proportional to
the number of templates, these rather unrealistic systems occupy quite a large
area and by excluding them in a search we can save significantly on the number
of templates.}
\label{fig:parameterspace}
\end{figure}

It is not always intuitive to translate the parameter space
into number of templates by looking only at the mass range. While
scaling-laws for the number of templates in terms of chirp times 
and lower-frequency cutoff do exist \cite{OwenSathyaprakash98},
the boundary effects are far too important when the parameter space
volume is not too large compared to the volume coverage of each
template. This is especially important towards the lower-end of the
mass range where the two curves given by $\eta=1/4$ and $m_1=m_{min}$ 
(cf. dashed and solid curves in Fig.~\ref{fig:t0t3}), 
meet and the template space is almost
one-dimensional. For instance, extending the mass
range of black holes from 30 $M_\odot$ to say 40 $M_\odot$ does not involve a
significant increase in the number of templates. On the other hand, decreasing the lower mass
of the binary PBH search has a deep implication: in the case of LIGO-I design sensitivity
curve, the BBH mass range [3-30] $M_\odot$ is covered with 774 templates; extending
the range to 40 $M_\odot$ changes the number of templates to 851. 
In the case of binary PBHs, going down from
0.3 $M_\odot$ to 0.1 $M_\odot$ increases the number of templates
by a factor of 30. In Fig.~\ref{fig:parameterspace} we have shown
the mass range for different astrophysical sources (LIGO PSD) and the
number of template needed to have a 95\% minimal match. The number of templates
needed to cover the BBH area is of the order of a thousand. BNS and BHNS is
four times more. And the PBHs case implies ten times more templates than the
BNS case. 

We can extend the parameter to any possibility including hypothetical cases such
as a binary composed of a PBH and a neutron star. The squares show binaries made
of similar compact objects (BBH, BNS, binary PBHs) whereas the rectangles 
show cases where the component stars are different such as BH-NS, NS-PBH and
BH-PBH binaries.  The following simulations do not include the last two cases
but we believe that our bank will work for these cases as well. 

\subsubsection{BNS case}
\begin{figure}[tbh]
\centering
\includegraphics[width=0.45\textwidth,angle=0]
 {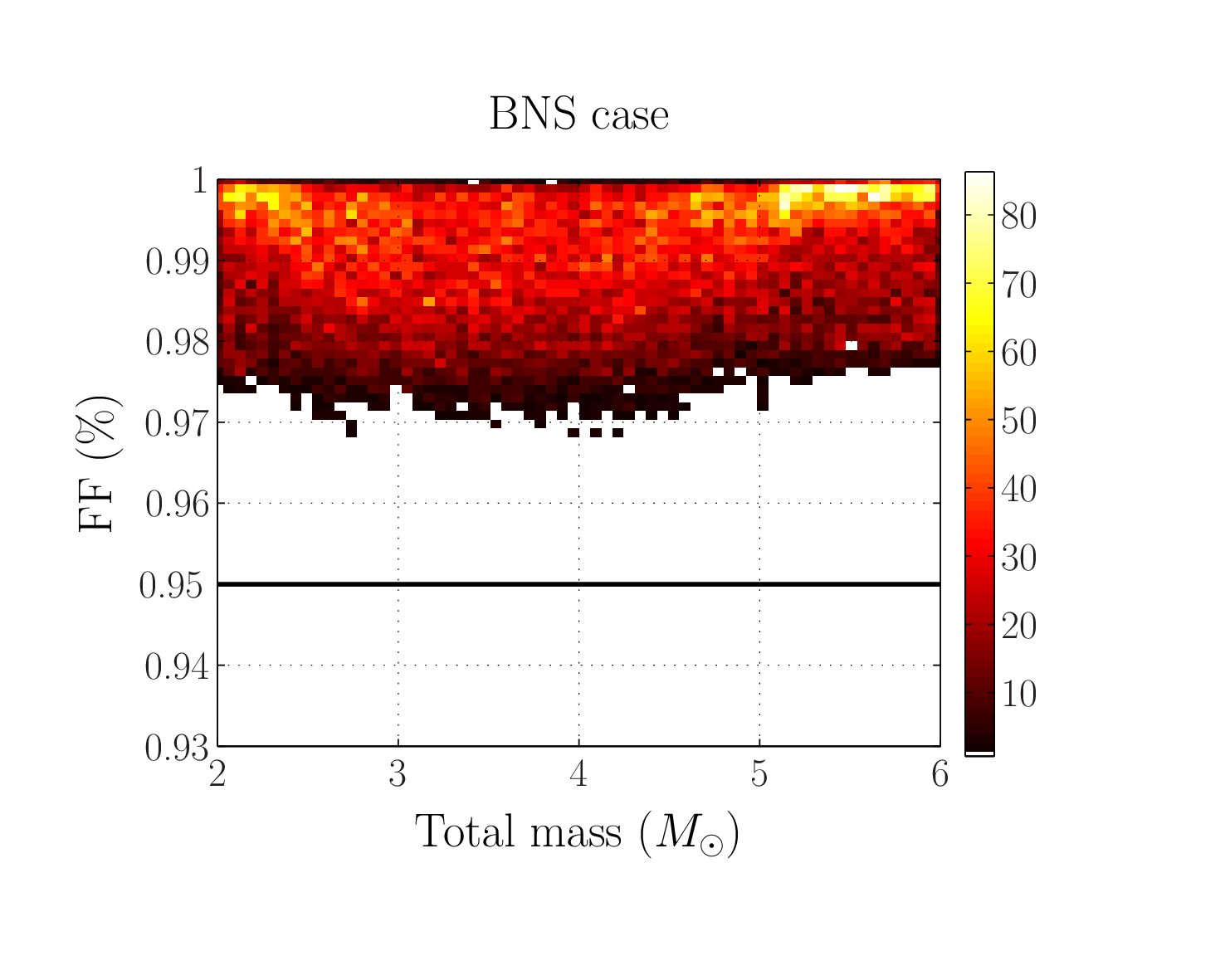}
\includegraphics[width=0.45\textwidth,angle=0]
{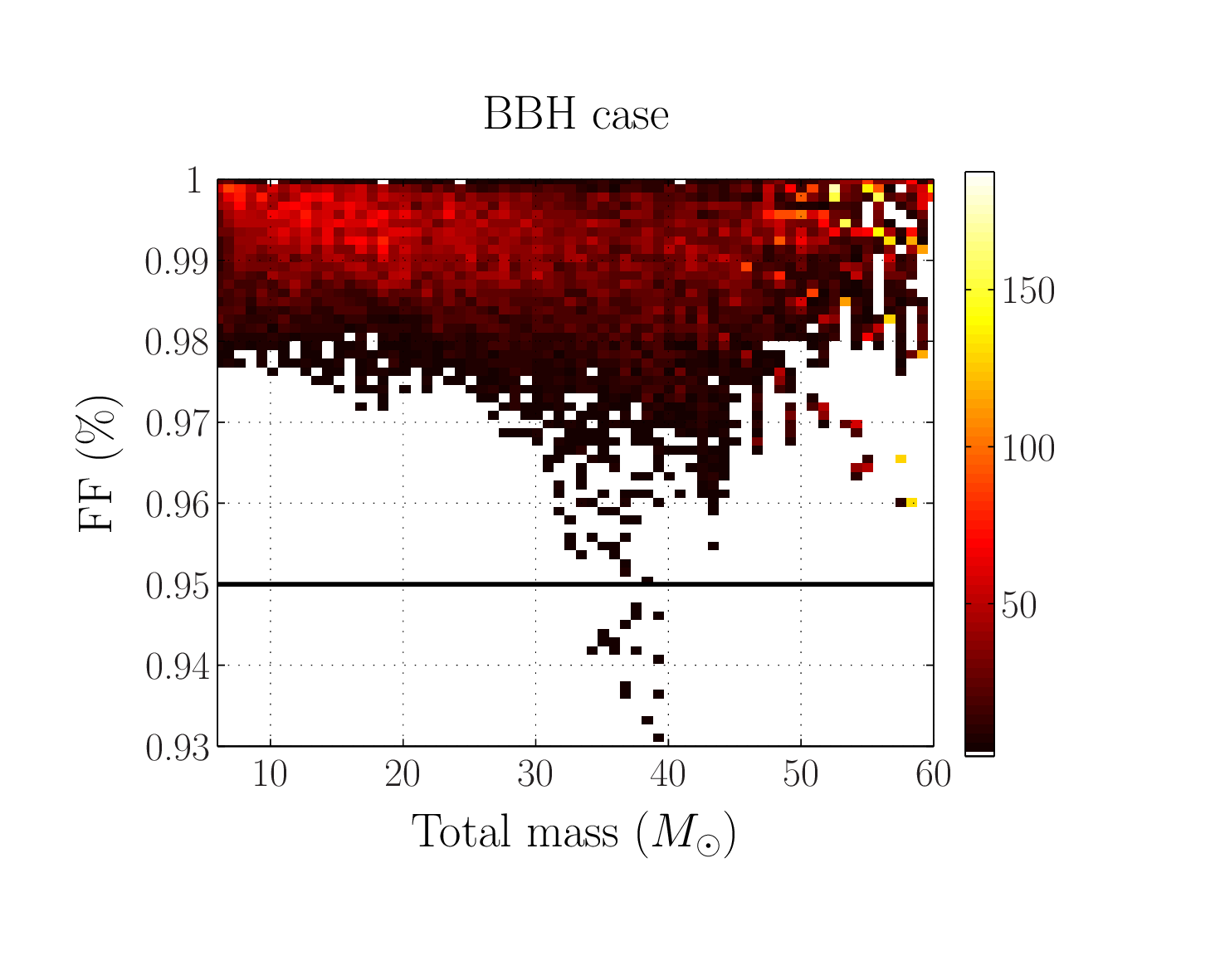}
\includegraphics[width=0.45\textwidth,angle=0]
{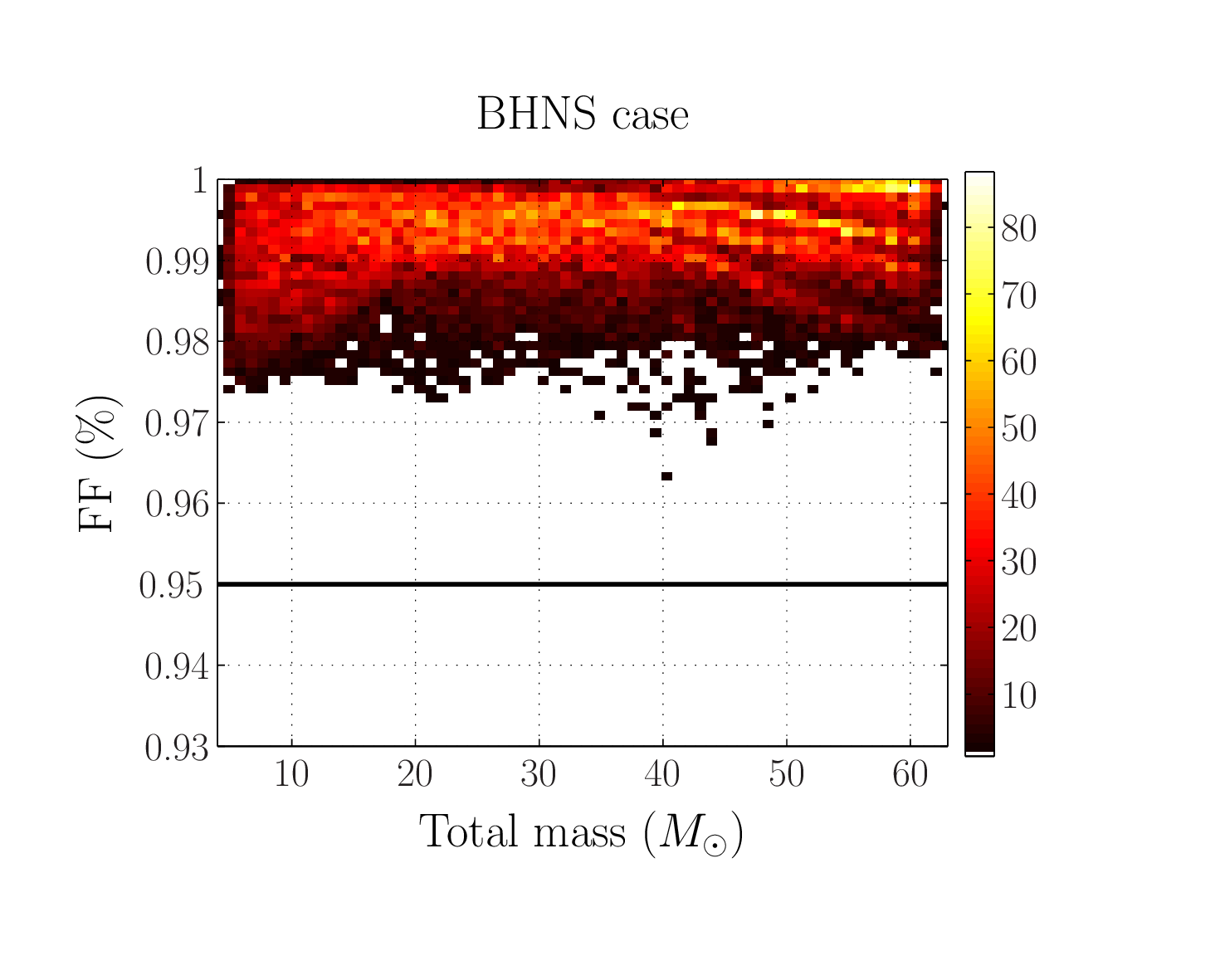}
\includegraphics[width=0.45\textwidth,angle=0]
{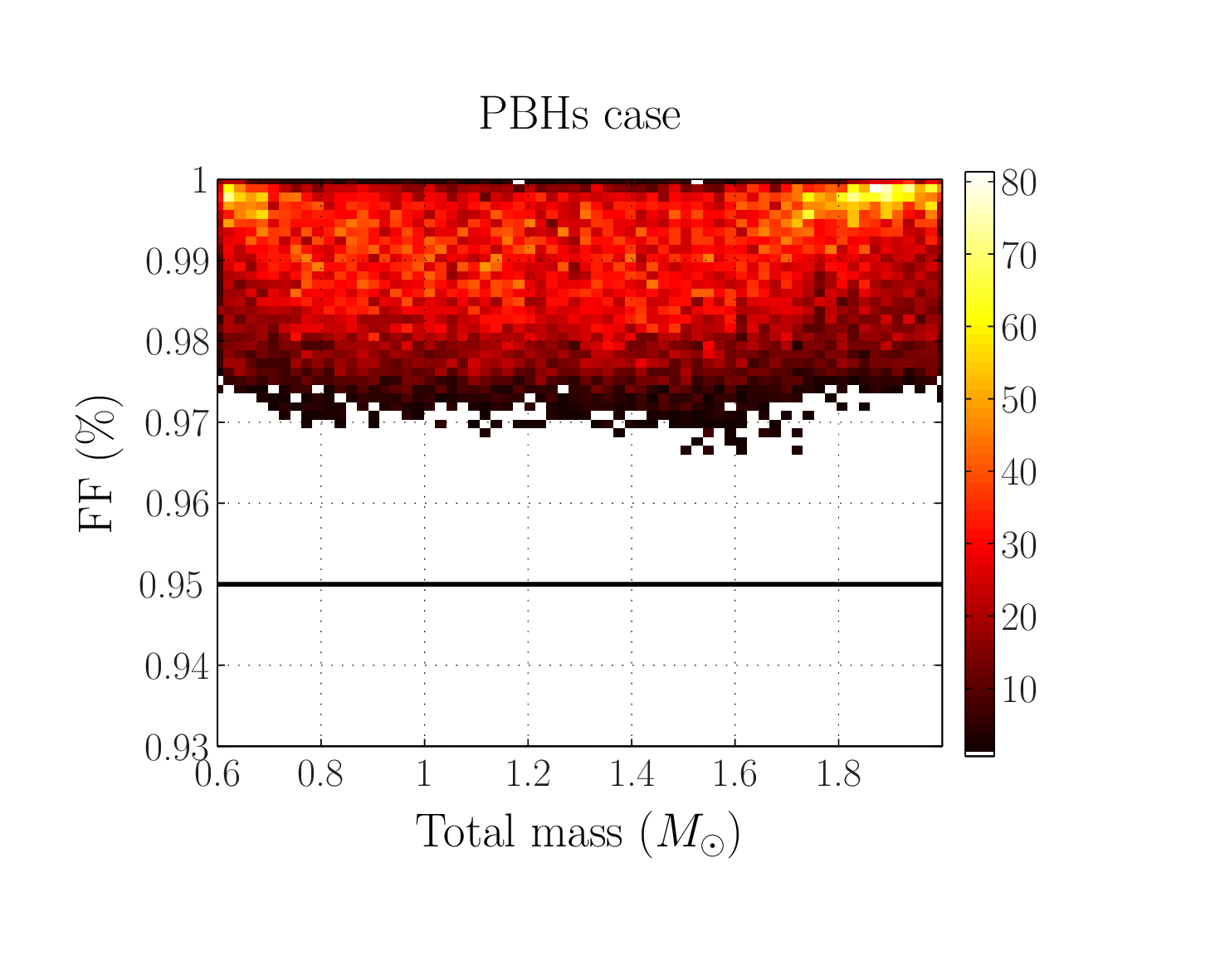}
\caption{Distribution of the efficiency
vectors versus total mass. The color bar on the right of each plot
shows the number of recovered signals with given values of fitting 
factor and total mass.}
\label{fig:efficiencyMhisto} 
\end{figure}

First, we focus on the BNS case in which the mass range is set to
1-3 $M_\odot$.  We plot
bank efficiencies versus the total mass in
Fig.~\ref{fig:efficiencyMhisto}. In these plots, we have combined the results of
the 4 simulations involving GEO\,600, LIGO-I, advanced LIGO and Virgo. In
Fig.~\ref{fig:cumFF}, we have separated the 4 results by plotting the
cumulative histogram of the efficiencies. We can see that in all cases
the safeness is above the minimal match:
$\mathcal{S}_{10^4}=0.966,$ therefore the bank is efficient for the search of
BNS. Furthermore, the cumulative histogram drops quickly while reaching the
safeness. This is an indicator of the good behaviour of the bank. This 
also indicates that the bank is over-efficient and it is partly due
to our choice of a rectangular, as opposed to a hexagonal, grid
and partly due to the extra templates along the equal-mass curve.

\subsubsection{Binary PBHs}
The main difference compared to the BNS case is the number of templates.
Indeed, we restrict the mass range to $[0.3-1]M_\odot$ and, therefore, the
number of
templates increases by a factor of 10. 
Results are summarised in Fig.~\ref{fig:efficiencyMhisto} and
Fig.~\ref{fig:cumFF}. The conclusions are very similar to the BNS case. The
safeness parameter is $\mathcal{S}_{10^4}=0.962$ and therefore the bank is
efficient for the search of binary PBHs.

\subsubsection{BBH case}\label{subsec:bbh}
The BBH search requires far fewer templates than in the case of binary PBHs 
or BNS.  However, it is important to recall that the duration of the signal 
decreases as the chirp mass increases and there is a critical mass above
which no signal enters the detector bandwidth. In the following
simulations we restrict the upper mass to be 30 $M_\odot$. 

\begin{figure}[tbh]
\centering
\includegraphics[width=0.45\textwidth,angle=0]
 {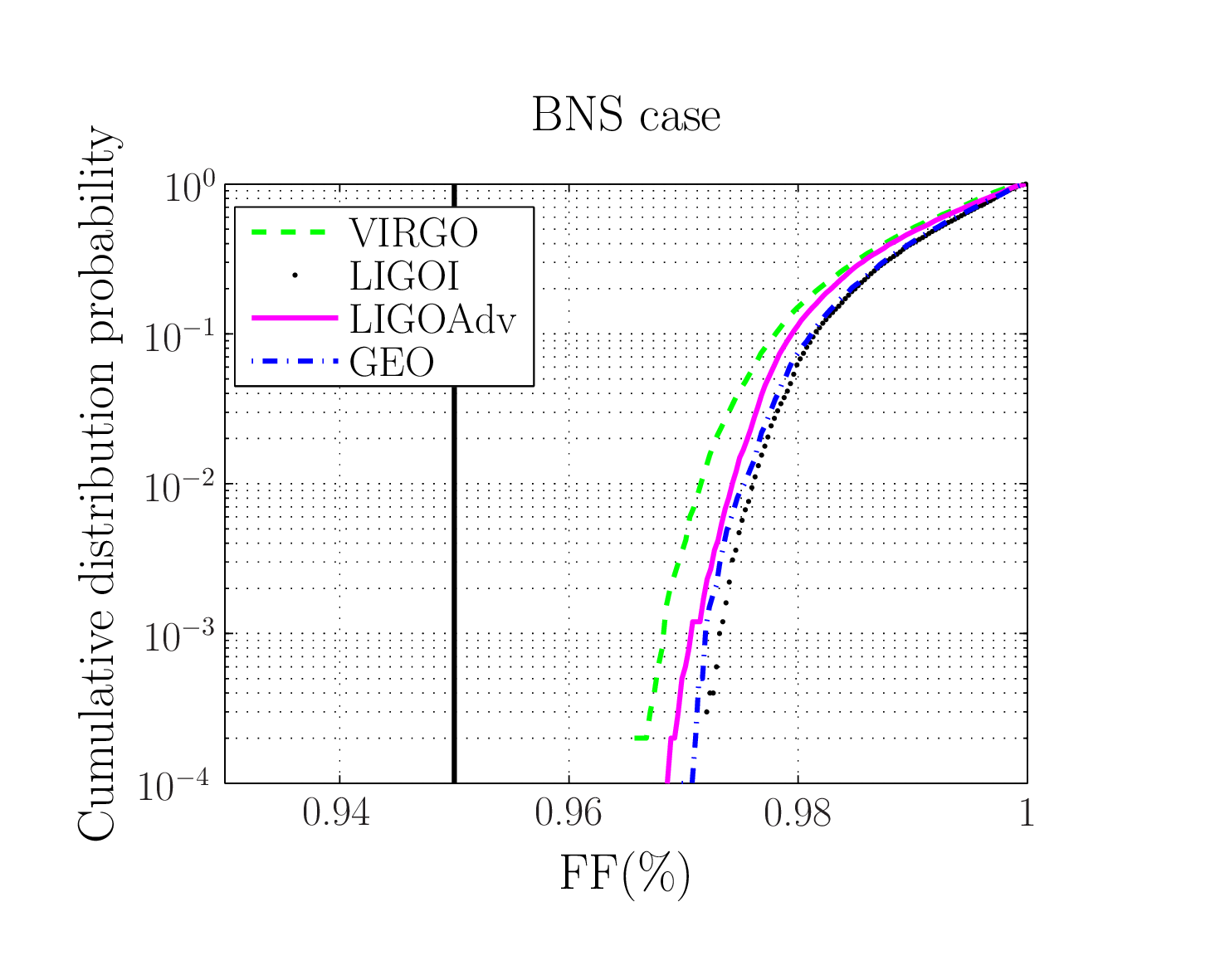}
\includegraphics[width=0.45\textwidth,angle=0]
{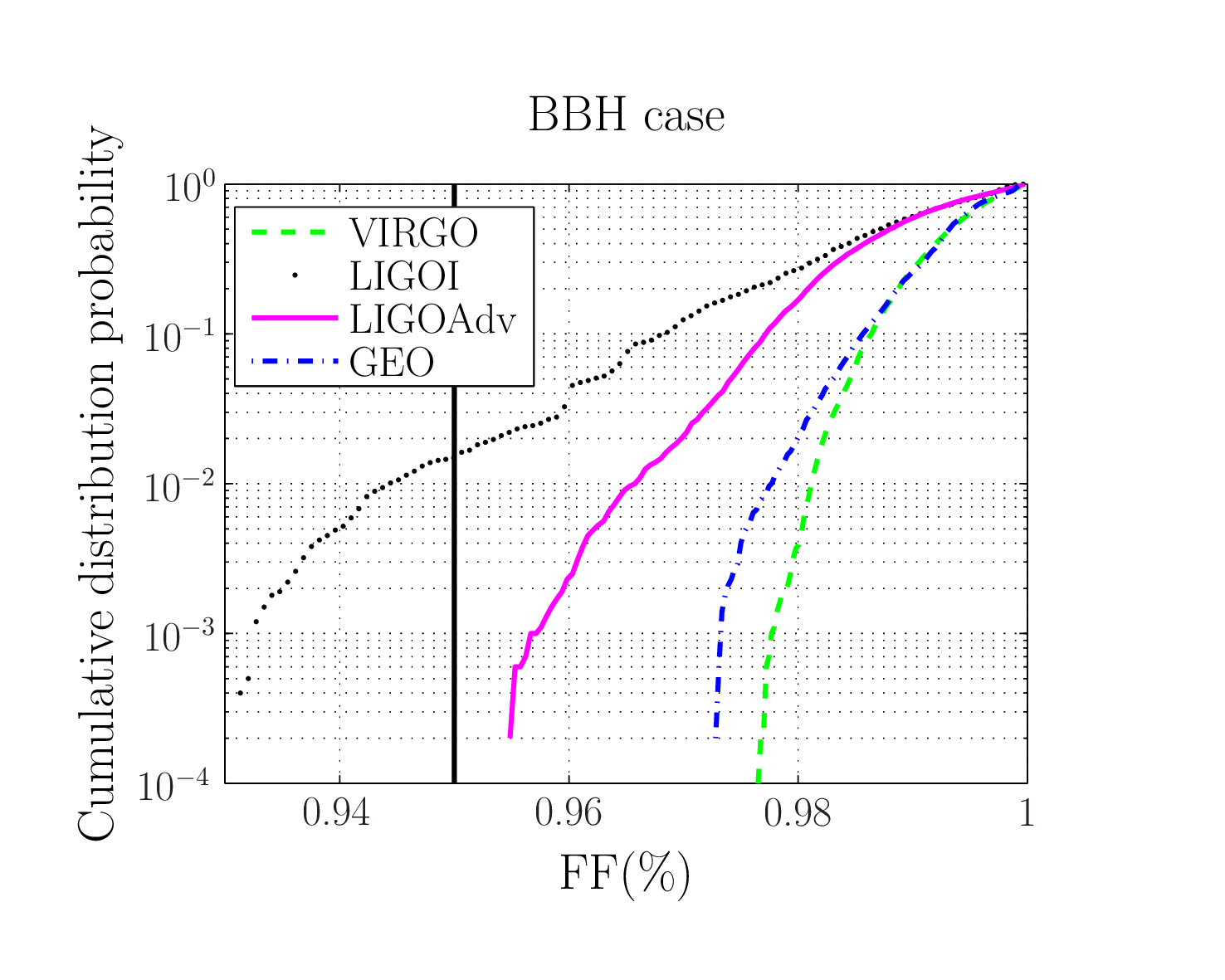}
\includegraphics[width=0.45\textwidth,angle=0]
{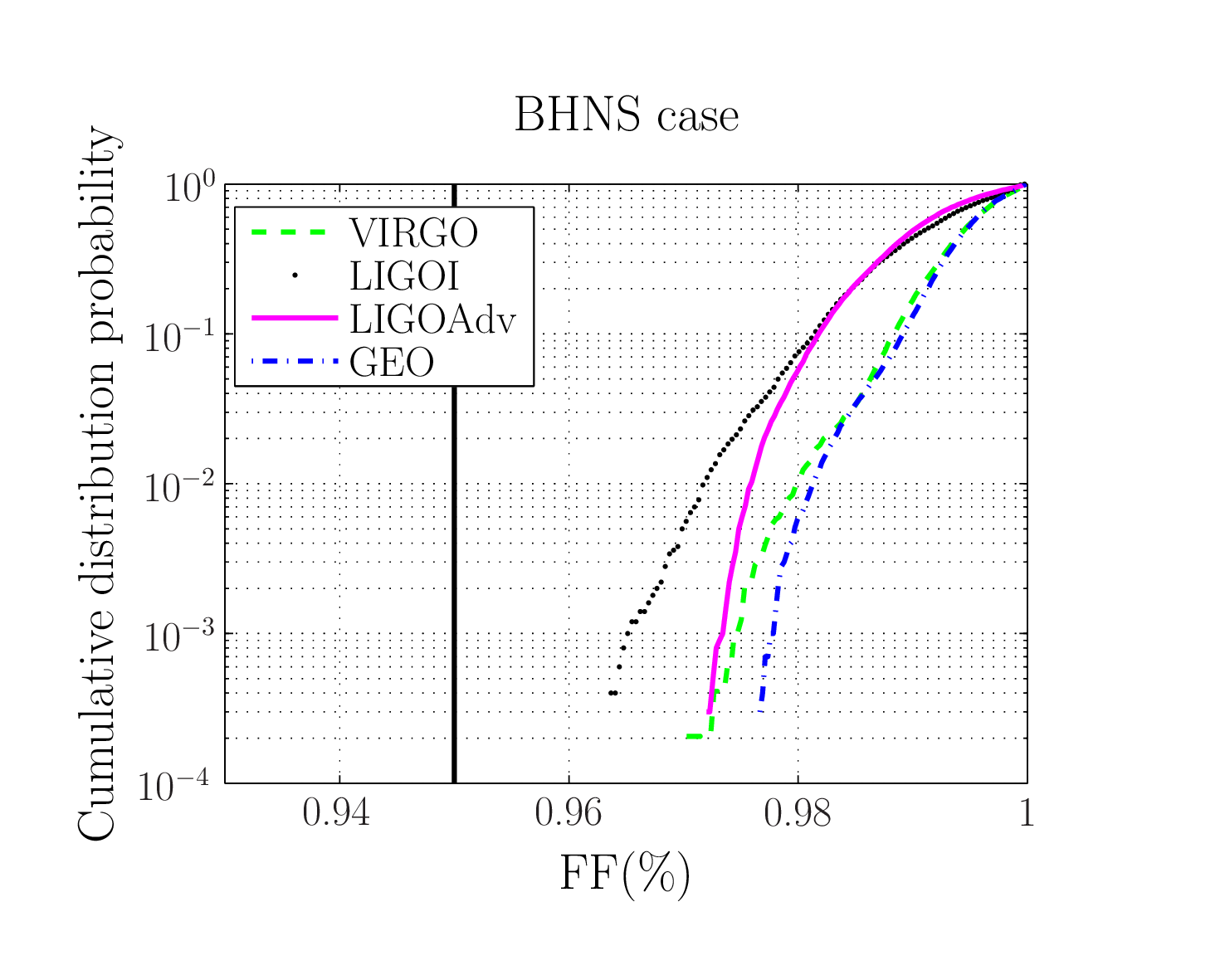}
\includegraphics[width=0.45\textwidth,angle=0]
{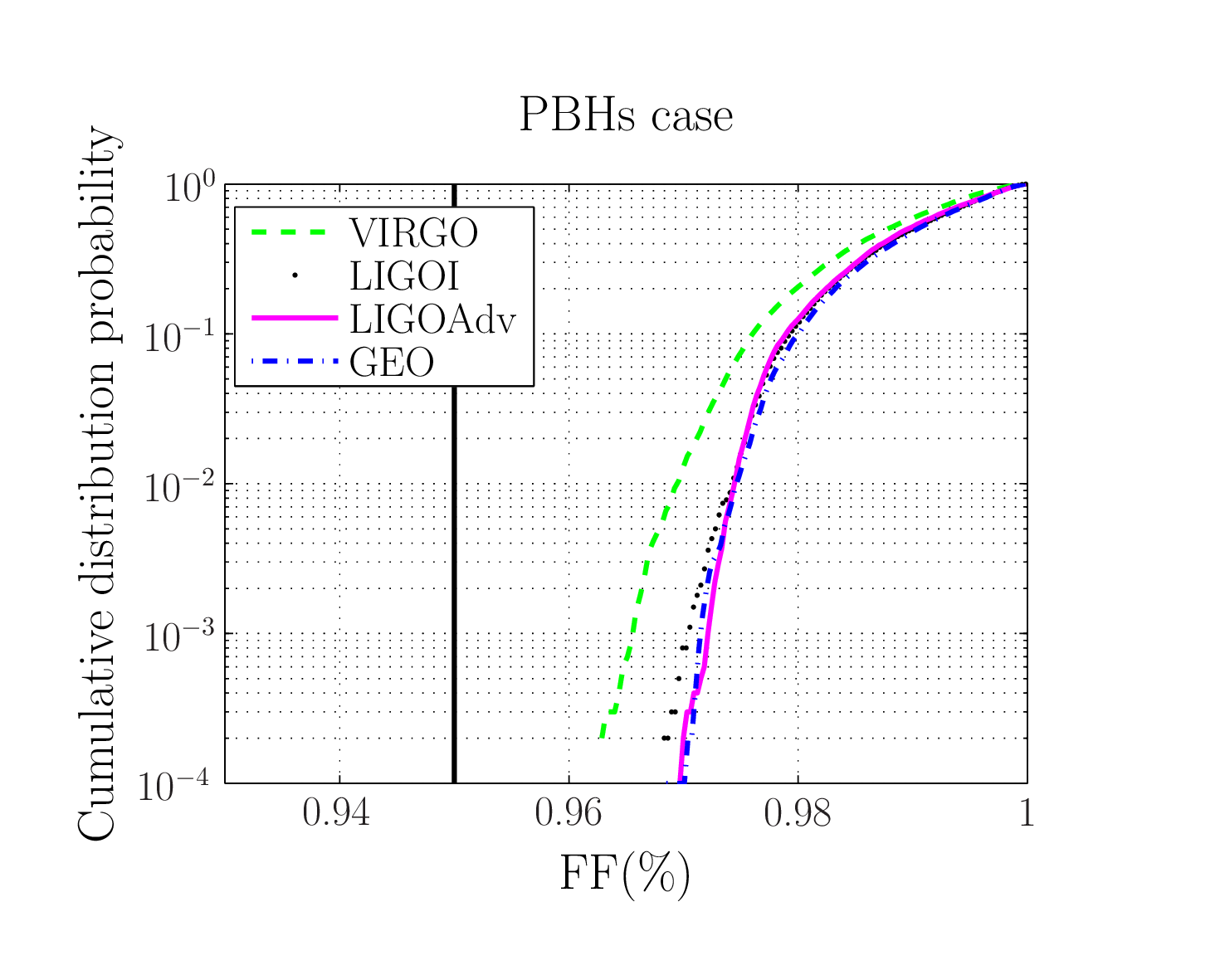}
\caption{Panels from top-left to bottom-right summarize the
cumulative distribution probability of the efficiency vector $\mathcal{E}$
for the different ranges of masses considered (namely, BNS, BBH, BH-NS and
binary PBHs). Each panel has results from all the different detectors.
While BNS, BH-NS and binary PBHs show a quadratic drop when FF drops closer to
the minimal match, the BBH case shows a slower drop in the case of LIGO-I and
advanced LIGO, with some values below the minimal match.}
\label{fig:cumFF} 
\end{figure}

\begin{figure}[tbh]
\centering
\includegraphics[width=0.45\textwidth,angle=0] {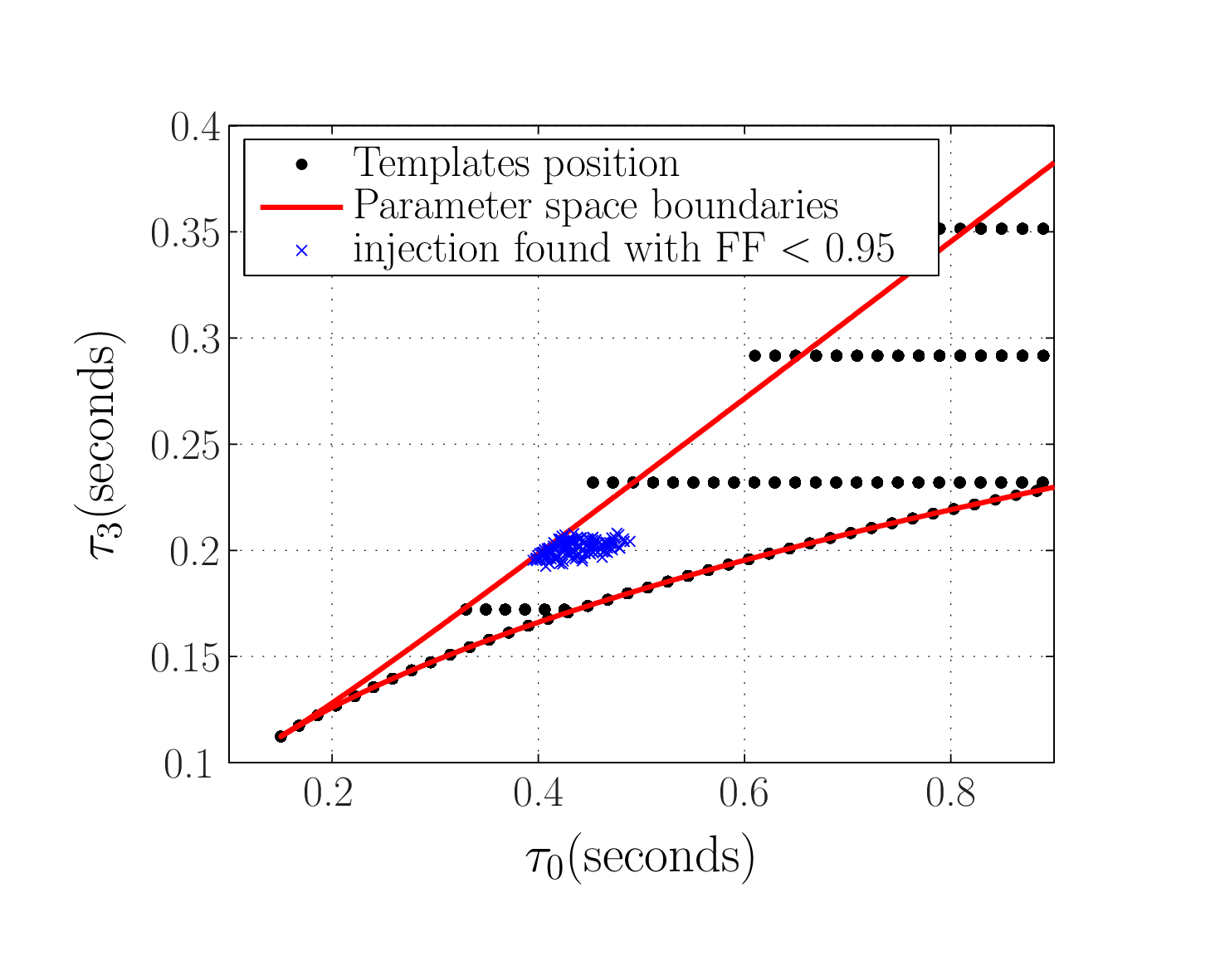}
\caption{The bank designed for the LIGO-I design sensitivity 
curve covers the mass range [3-30] $M_\odot,$ but there is 
a `hole' in the parameter space corresponding to the total mass 
in the range $[30,\, 40] M_\odot$. Black dots are where the templates
are placed and crosses are the positions of the injected signals
that have a fitting factor below the minimal match; majority 
of the injections are found with $FF>MM$ and are not plotted 
for the sake of clarity.}
\label{fig:hole}
\end{figure}

By studying the results in Fig.~\ref{fig:efficiencyMhisto} and \ref{fig:cumFF} and 
Table~\ref{tab:safeness}, it is clear that the bank is
efficient in the case of GEO\,600, advanced LIGO and Virgo detectors, with the
safeness equal to 0.973, 0.955 and 0.976, respectively, but in the 
LIGO-I case $\mathcal{S} < MM.$  From Fig.~\ref{fig:efficiencyMhisto} we
see that there are fewer than 1\% of the injections that have a fitting factor
below $MM.$ However, these few injections do have an overlap larger than
0.930. The injections that fail to achieve the required match
are all concentrated in a specific region of the parameter space with
their total mass between 30 and 40 $M_\odot$ as seen from Fig.~\ref{fig:hole}.
This is the region where
the approximation that the signal manifold is flat breaks down and causes
a ``small hole'' in template placement; but the effect is insignificant
and does not deserve a fix. In any case, it is possible to use a larger
$MM$ of the bank so as to capture these signals with the desired match.
The increase in the number of templates is not a major issue since the 
number of templates in the BBH search is quite small
compared to binary PBHs or BNS.
\begin{table}[b]
\caption{Safeness parameter derived from the bank simulations.}
\label{tab:safeness}
\begin{ruledtabular}
\begin{tabular}{cccccc}
Safeness $\mathcal{S}_{10^4}$ versus detector & GEO\,600 & LIGO-I & Advanced
LIGO&
VIRGO & all\\\hline
BBH& 0.973&    0.931&    0.955&    0.976& 0.931\\
BH-NS&0.977&    0.964&    0.972&    0.970&0.964\\
BNS&0.970&0.972&0.968&0.966&0.966\\
Binary PBHs&    0.968&    0.968&    0.969&    0.962&0.962\\
\end{tabular}
\end{ruledtabular}
\end{table}

\subsubsection{BH-NS case}
Finally, we look at the BH-NS case. Here the component masses 
can take values in the range $[1-30]M_\odot$. There is an 
important issue here related to both the injections
and ranges of masses in the template bank. First, since the BNS and BBH cases have 
already been studied independently, we do not need to look again at  
BBH or BNS systems. Therefore, we restrict $m_1 \in [1-3]M_\odot$ and 
$m_2 \in [3-60]M_\odot$. Second, the bank 
has to be extended up to 63 $M_\odot,$ which is the sum of the maximum mass of
the two objects.  The safeness in this case is
above the minimal match with a value $\mathcal{S}_{10^4}=0.964,$ which means that
the proposed bank is efficient for the BH-NS search as well.

\subsection{Taylor models}
\begin{figure}[tbh]
\centering
\includegraphics[width=0.45\textwidth,angle=0]
 {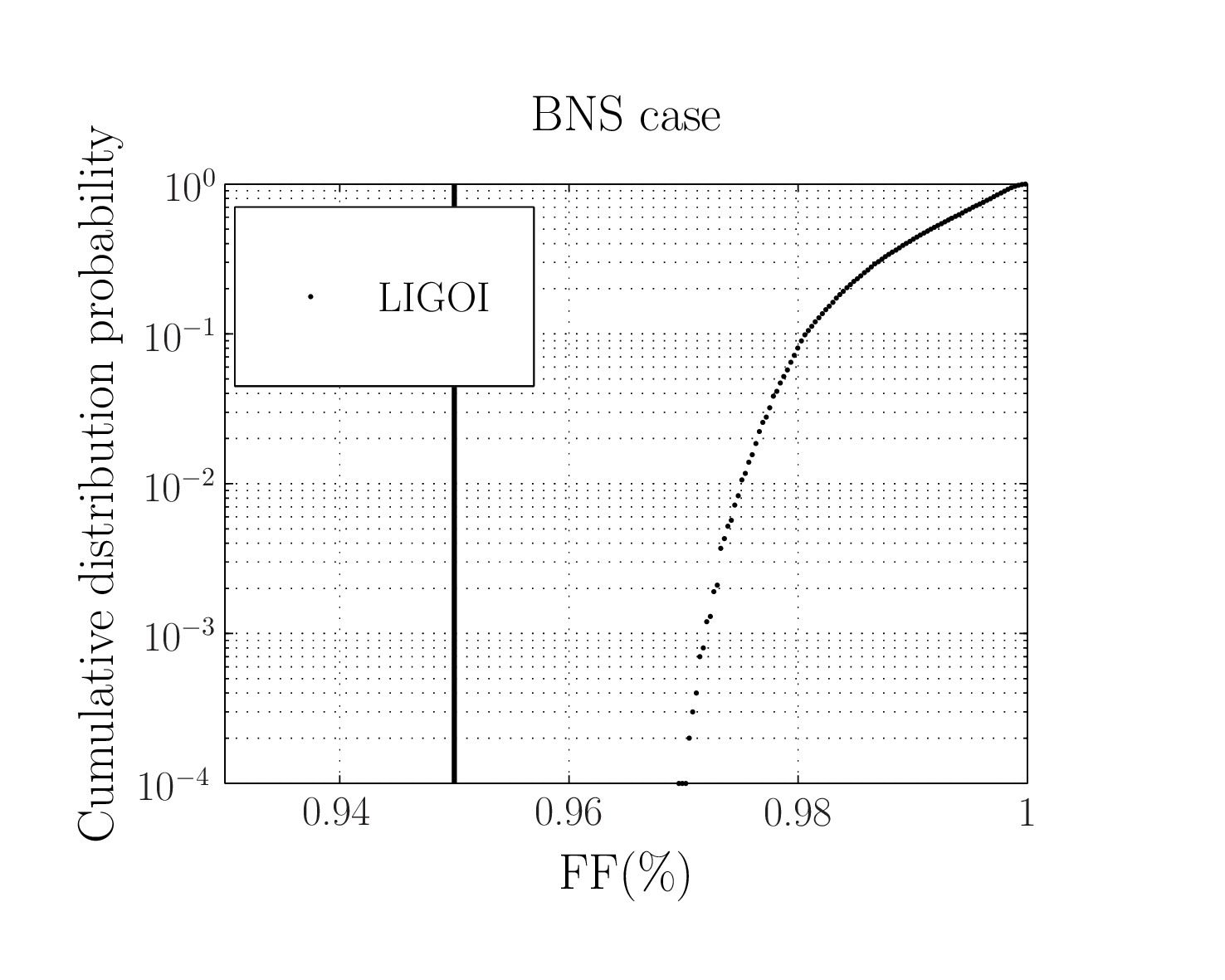}
\includegraphics[width=0.45\textwidth,angle=0]
{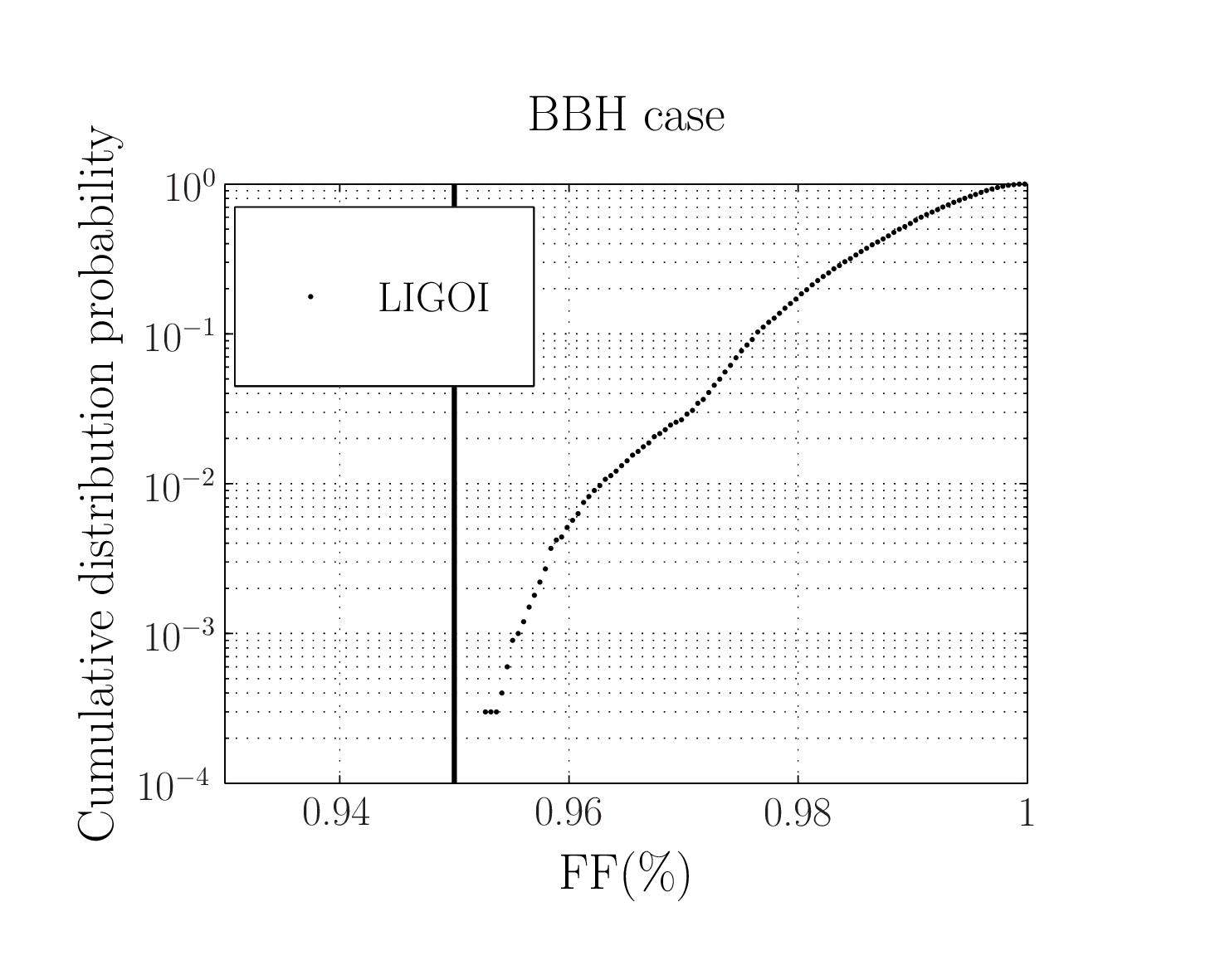}
\includegraphics[width=0.45\textwidth,angle=0]
{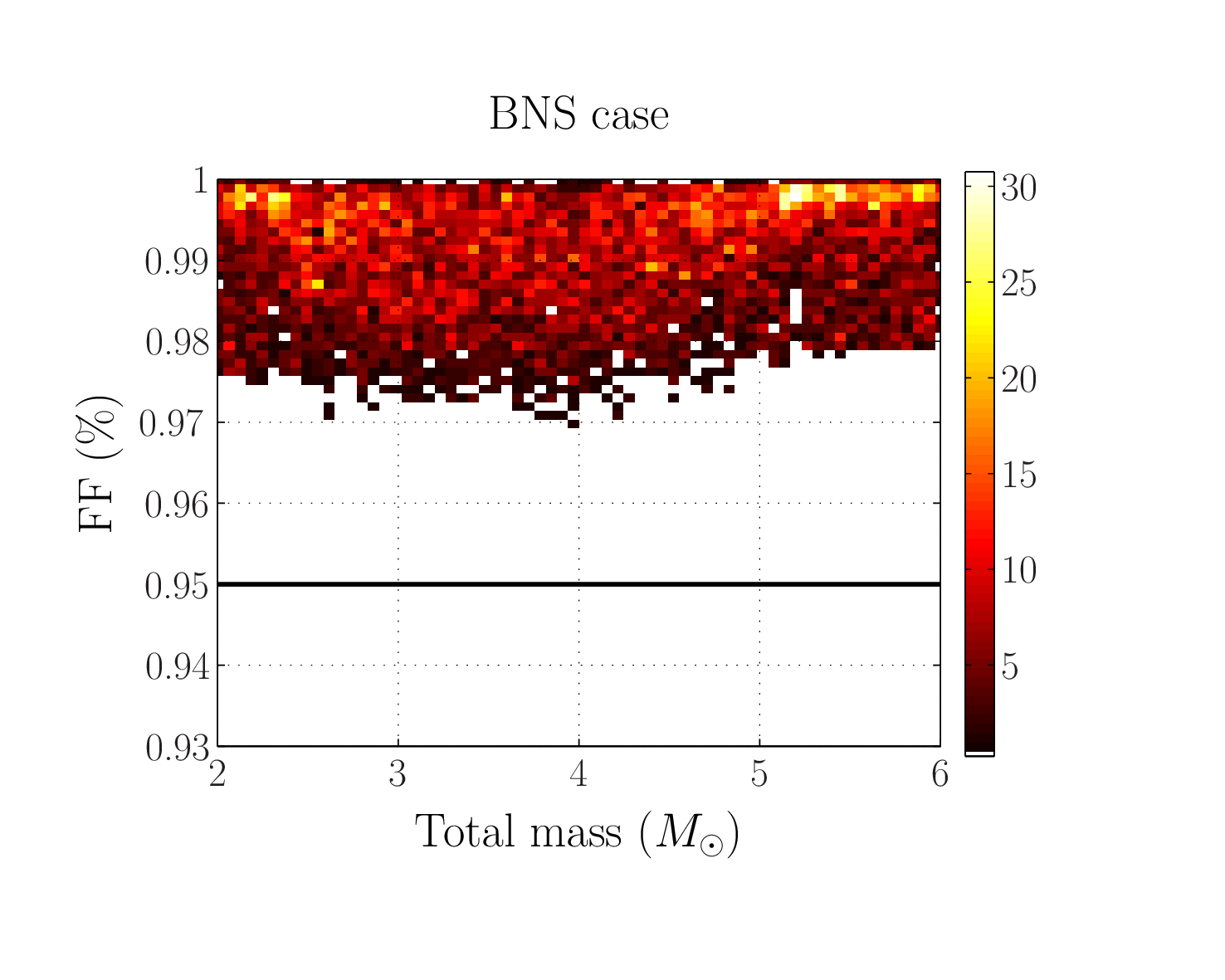}
\includegraphics[width=0.45\textwidth,angle=0]
{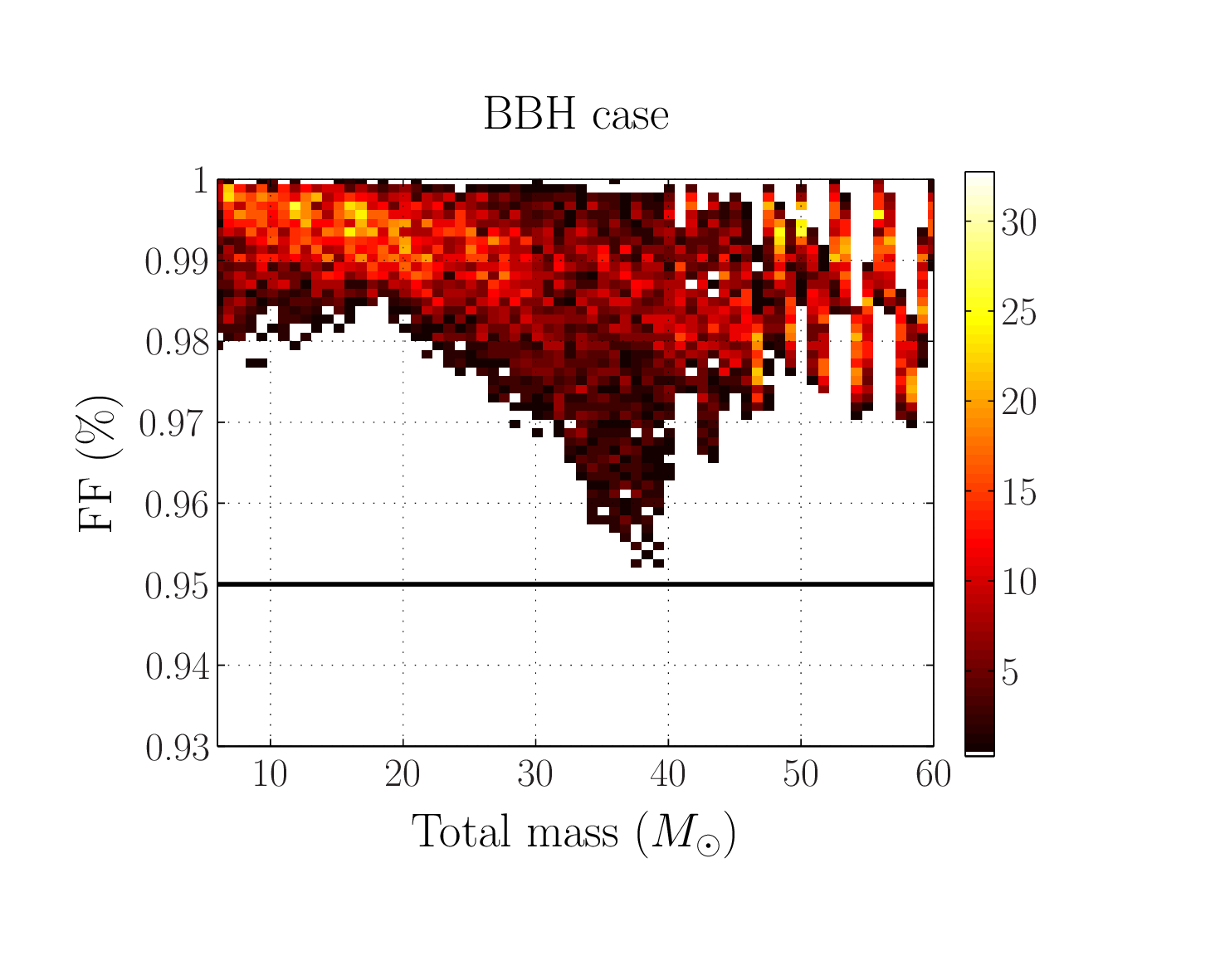}
\caption{Cumulative distribution probability of the efficiency vectors
$\mathcal{E}$ in the
LIGO-I design sensitivity curve using a Taylor model instead of SPA both
for templates and injected signals (top figures) and $\mathcal{E}$ distribution
versus the total mass (bottom).}
\label{fig:efficiencyMhisto2} 
\end{figure}

Finally, we perform two simulations, in the BBH and BNS case using only LIGO-I
design sensitivity curve, in which both the templates and signals are 
based on the standard time-domain post-Newtonian model at 2PN order \cite{DIS2}. 
The parameters of the simulation are exactly the same as above. 
The results are shown in Fig.~\ref{fig:efficiencyMhisto2}.
As one can see, the recovered fitting factors are all above the $MM$ for both
BNS and BBH, which means that the bank designed in this
paper using the SPA models can be used for physical models based on Taylor
approximants as well.

\subsection{Computational cost}
The algorithm presented in this paper is relatively fast. The computation time
required to generate a template bank is of the order of a second
to a few tens of seconds on any standard computer, depending on the parameters
considered. Table \ref{tab:compcost}
summarizes the computational cost needed to generate different template banks
on a Linux Atlhon 1.6~GHz machine. The parameters used in these examples are 
similar to those in Ref.\ \cite{beauville}, that is $F_L=30$~Hz,
$f_s=4096$~Hz and $MM=0.95$ and a VIRGO PSD. The Table shows that a BBH
template bank is created in a second while a template bank with
$m_{\textrm{min}}=0.5M_\odot $ and $m_{\textrm{max}}=30M_\odot $ (more than 180,000 templates)
required only 28 seconds. The number of templates quoted in
Table \ref{tab:compcost} and in Ref.\ \cite{beauville} are in very good agreement if we
take into account a 30\% effect coming from the fact that we used a
square lattice instead of an hexagonal one.

Let us now consider our BBH template bank. The cost of 
matched filtering (to process 128 seconds of data through 2422 templates) is of
the order of 50 minutes while the template bank generation costs 1 second. It is
obvious that the template bank generation is a small fraction of the total
cost. 
\begin{table}[b]
\caption{Computational cost for different template banks.}
\label{tab:compcost}
\begin{ruledtabular}
\begin{tabular}{cccc}
$m_{\textrm{min}}$ & $m_{\textrm{max}}$ & N & Time(s)\\\hline
0.5 & 30 &  182137 & 28.0 \\
1   & 3  &  10188 & 4.5 \\
1   & 30 &  34905 & 6.0 \\
3   & 30 &   2422 & 1.0 \\
\end{tabular}
\end{ruledtabular}
\end{table}
Since the placement uses a square lattice the number of templates is expected to
be an extra 30\% more in comparison to an hexagonal grid and
the computational cost is larger by the same factor. 
Nowadays, analysing a month of gravitational wave data through $\approx 2000$
templates take a couple of days only depending on the number of computers
used. 

Decreasing the number of templates is, therefore, interesting  especially in the
BNS and PBH cases where the number of templates is large although this is not
a major issue in the BBH case where it might even be safer to use a square
lattice (see Sec.~\ref{subsec:bbh}). We provide an improvement of the
algorithm presented in this paper by incorporating an hexagonal
placement in a future work \cite{BCS2}.

\section{conclusion}\label{sec:conclusions}
In this paper we have used a geometrical method for constructing 
a template bank to search for gravitational waves from binary coalescences. 
The bank design is based on the Fourier-domain representation of the expected 
signal and the templates are placed on a square lattice. We have shown that 
the bank performs well with respect to its efficiency and safeness in the 
case of BNS, BH-NS and binary PBHs and for the 4 detectors considered and 
even over-efficient as expected since a square lattice has been used.  

The proposed bank is also efficient in the case of BBH when the total mass 
is in the range [6-30] $M_\odot$. It remains efficient above
30 $M_\odot$ up to 60 in the  case of VIRGO, advanced LIGO and GEO\,600
detectors, 
the only problem  being the case of LIGO-I in a specific total mass range of 
[30-40] $M_\odot$. This particular case has been investigated and correspond
to the  largest masses considered where the length of the signal is very short. 
Increasing the minimal match from 0.95 to 0.97 in the area which gives low
match cures the problem. Therefore, we conclude that the square bank 
proposed is efficient for all the PSDs and searches we studied in this paper.
We remind here that this bank has been used by the LIGO Scientific 
Collaboration to search for binary PBHs and BNS in four science
runs \cite{LIGOS2bns,LIGOS2bbh,LIGOS2PBH}. 

Finally, it is important to emphasize that though we have been using 
signals and templates based on the SPA model, there is no restriction 
in using the same bank but for different models of signals and templates. 
Indeed, we have found that the current bank is
efficient even when the signals and templates are both based on Taylor, 
Pad\'e or EOB models (see Ref.\ \cite{DIS2} for model classification) as shown in a 
companion paper \cite{BCS2}; in this paper we have also obtained a fast algorithm to
construct a template bank on an hexagonal lattice with the number of templates reduced by
$\approx 30\%$ (useful for the case of BNS and PBH searches where
the number of templates could be very large).

\appendix\label{sec:app}
\section{Placement algorithm}

In this appendix we provide the algorithm used for the template placement
described in Sec.~\ref{sec:template placement}. The algorithm is split into
two distinct parts depending on the position in the parameter space. 

In a nut-shell, the algorithm to lay templates along the equal-mass curve is as
follows:
\begin{obeylines}
\texttt{
\hskip 1 true cm Begin at $\tau_0 = \tau_0^{\rm min}$ 
\hskip 1 true cm do while $(\tau_0 < \tau_0^{\rm max})$
\hskip 1 true cm \{
\hskip 2 true cm $\tau_0^A = \tau_0 + \delta\tau_0, \ \ \tau_3^A = 4A_3\left({\tau_0^A}/{4A_0} \right)^{2/5}$ 
\hskip 2 true cm $\tau_3^B = \tau_3 + \delta\tau_3, \ \ \tau_0^B = 4A_0 \left({\tau_3^B}/{4A_3} \right)^{5/2}$ 
\hskip 2 true cm if ($(\tau_0^A,\tau_3^A)$ is closer to $(\tau_0,\tau_3)$ than $(\tau_0^B,\tau_3^B)$)
\hskip 2 true cm \{
\hskip 3 true cm $\tau_0 = \tau_0^A, \tau_3 = \tau_3^A$ 
\hskip 2 true cm \}
\hskip 2 true cm else
\hskip 2 true cm \{
\hskip 3 true cm $\tau_0 = \tau_0^B, \tau_3 = \tau_3^B$ 
\hskip 2 true cm \}
\hskip 2 true cm Add $(\tau_0, \tau_3)$ to InspiralTemplateList
\hskip 2 true cm numTemplates++
\hskip 2 true cm Compute metric at $(\tau_0, \tau_3)$
\hskip 2 true cm Compute distance between templates at  new point:$(\delta\tau_0, \delta\tau_3)$ 
\hskip 1 true cm \}
}
\end{obeylines}

The algorithm to lay templates in the rest of the parameter space is as follows:
\begin{obeylines}
\texttt{
\hskip 1 true cm Begin at $\tau_0 = \tau_0^{\rm min}, \tau_3 = \tau_3^{\rm min}$ 
\hskip 1 true cm Compute metric at $(\tau_0, \tau_3)$
\hskip 1 true cm Compute distance between templates at  new point:$(\delta\tau_0, \delta\tau_3)$
\hskip 1 true cm Add $(\tau_0, \tau_3)$ to InspiralTemplateList
\hskip 1 true cm numTemplates++
\hskip 1 true cm do while ($\tau_3 <= \tau_3^{\rm max}$)
\hskip 1 true cm \{
\hskip 2 true cm Begin at $\tau_0 = \tau_0^{\rm min}$ 
\hskip 2 true cm do while ($\tau_0 <= \tau_0^{\rm max}$)
\hskip 2 true cm \{
\hskip 3 true cm if ($(\tau_0, \tau_3)$ is inside the parameter space)
\hskip 3 true cm \{
\hskip 4 true cm Compute metric at ($\tau_0, \tau_3$)
\hskip 4 true cm Compute distance between templates at  new point:($\delta\tau_0, \delta\tau_3$) 
\hskip 4 true cm Add ($\tau_0, \tau_3$) to InspiralTemplateList
\hskip 4 true cm numTemplates++
\hskip 3 true cm \}
\hskip 3 true cm Increment $\tau_0:$ $\tau_0 = \tau_0 + \delta\tau_0$ 
\hskip 2 true cm \}
\hskip 2 true cm Increment $\tau_3:$ $\tau_3 = \tau_3 + \delta\tau_3$ 
\hskip 2 true cm Get next template along $\tau_3={\rm const.}$:$(\tau_0,\tau_3)$
\hskip 1 true cm \}
}
\end{obeylines}
This algorithm is very simple to implement: except for the metric computation,
the code is based on loops over simple calculations. It is a fast and
robust algorithm amenable to easy implementation.

\begin{acknowledgments}
This research was supported partly by Particle Physics and Astronomy
Research Council, UK, grant PP/B500731.
The authors benefitted from useful discussions with the members of 
the LIGO Scientific Collaboration, in particular B.\ Allen, P.\ Brady, 
D.\ Brown, J.\ Creighton and B.J.\ Owen.
\end{acknowledgments}

\label{theend}
\end{document}